\documentclass[superscriptaddress,prd,preprint,nofootinbib]{revtex4}
\usepackage{amsmath,amssymb}
\usepackage{graphicx,epsfig}

\def\simgt{\mathrel{\lower2.5pt\vbox{\lineskip=0pt\baselineskip=0pt
           \hbox{$>$}\hbox{$\sim$}}}}
\def\simlt{\mathrel{\lower2.5pt\vbox{\lineskip=0pt\baselineskip=0pt
           \hbox{$<$}\hbox{$\sim$}}}}

\def\vev#1{\left\langle#1\right\rangle}
\def\beq{\begin{equation}}
\def\eeq{\end{equation}}
\def\beqa{\begin{eqnarray}}
\def\eeqa{\end{eqnarray}}
\newcommand{\gsim}{ \mathop{}_{\textstyle \sim}^{\textstyle >} }
\newcommand{\lsim}{ \mathop{}_{\textstyle \sim}^{\textstyle <} }

\newcommand{\Dsl}{\mbox{\ooalign{\hfil/\hfil\crcr$D$}}}

\newcommand{\KEV}{ {\rm keV} }
\newcommand{\MEV}{ {\rm MeV} }
\newcommand{\GEV}{ {\rm GeV} }
\newcommand{\TEV}{ {\rm TeV} }

\def\Re{\mathrm{Re}}

\begin{document}

\preprint{UCB-PTH-04/14}
\preprint{LBNL-55001}

\title{Electroweak Supersymmetry with An Approximate $U(1)_{PQ}$}

\author{Lawrence J.~Hall}

\affiliation{Department of Physics, University of California, Berkeley, and\\
      Theoretical Physics Group, Lawrence Berkeley National Laboratory,\\
      Berkeley, CA 94720, USA}

\author{Taizan Watari}
\affiliation{Department of Physics, University of California, Berkeley, and\\
      Theoretical Physics Group, Lawrence Berkeley National Laboratory,\\
      Berkeley, CA 94720, USA}

\begin{abstract}
A predictive framework for supersymmetry at the TeV scale is presented, 
which incorporates the Ciafaloni--Pomarol mechanism for the dynamical 
determination of the $\mu$ parameter.
The $\mu$ parameter of the MSSM is replaced by $\lambda S$,
where $S$ is a singlet field, and the axion becomes a heavy
pseudoscalar, $G$, by adding a mass, $m_G$, by hand. 
The explicit breaking of Peccei--Quinn (PQ) symmetry is assumed to be
sufficiently weak at the TeV scale that the only observable consequence 
is the mass $m_G$. Three models for the explicit PQ 
breaking are given; but the utility of this framework is that 
the predictions for all physics at the electroweak scale are
independent of the particular model for PQ breaking. 
This framework leads to a theory similar to the MSSM, except
that $\mu$ is predicted by the Ciafaloni--Pomarol relation, 
and there are light, weakly-coupled states that lie dominantly 
in the superfield $S$. The production and cascade decay of
superpartners at colliders occurs as in the MSSM, except that there is
one extra stage of the cascade chain, with the next-to-LSP decaying
to its ``superpartner'' and $\tilde{s}$, dramatically altering the collider
signatures for supersymmetry. The framework is compatible with terrestrial 
experiments and astrophysical observations for a wide range of $m_G$ and
$\vev{s}$. If $G$ is as light as possible, $300 \; \KEV < m_G < 3 \MEV$,
it can have interesting effects on the radiation energy density
during the cosmological eras of nucleosynthesis and acoustic
oscillations, leading to predictions for $N_{\nu BBN}$ and
$N_{\nu CMB}$ different from 3.  
\end{abstract}

\maketitle
\section{ Introduction}
\label{sec:intro}

If nature is supersymmetric at the TeV scale, the weakness of gravity 
can be naturally understood and a highly successful prediction for the
weak mixing angle results. For well over 20 years, theorists have
examined the possible forms of supersymmetric electroweak theories,
and experimentalists have pondered how superpartners may be discovered.
Although we talk of a minimal supersymmetric standard model, the MSSM, it
is not clear that this model is preferred over others. Below we
examine the stages required to supersymmetrize the known gauge
interactions, and argue for a new simple phenomenological framework.

The first stage is to place the known elementary particles into
multiplets of supersymmetry.
The quarks and leptons of the standard model $q,u,d,l,e$ are placed in
chiral multiplets $Q,U,D,L,E$, implying that spin zero squarks and
sleptons are expected at the weak scale, while the $QCD$ and
electroweak gauge bosons are placed in vector multiplets of
supersymmetry, leading to spin 1/2 gluinos, winos and photinos.
The next stage is to supersymmetrize the Higgs boson of the standard
model. If it is placed in a single chiral multiplet, the corresponding
Higgsino leads to gauge anomalies, and this single Higgs multiplet is
not able to give masses to both the up quarks and the down quarks. The
minimal possibility is two Higgs chiral multiplets $H_1,H_2$. The quark
and lepton interactions with the Higgs are described by the
superpotential
\begin{equation}
W = \lambda_U \; QU H_2 +  \lambda_D \; QD H_1 +  \lambda_E \; LE H_1.
\label{eq:W}
\end{equation}
The supersymmetric interactions contain both Yukawa couplings between
the fermions and scalars, and quartic interactions between the
scalars. Similarly, the supersymmetric gauge interactions involve
Yukawa and quartic type interactions, but no new parameters are needed
beyond those of the standard model.

Few would doubt that any supersymmetric electroweak theory in 4 
dimensions must contain this minimal set of interactions. The
real question is: what else is needed? The most glaring omission is
the breaking of supersymmetry. It is possible to remain
agnostic about the fundamental origin of supersymmetry breaking: 
one can simply assume that in the TeV scale effective theory
the breaking is described by a set of
operators that do not spoil the controlled radiative behaviour of
supersymmetric theories \cite{DG}. In practice this means that gaugino masses
and scalar mass terms can be added by hand, together with a certain
set of bilinear and trilinear scalar interactions, one for each term
in the superpotential. One can leave the
origin of these ``soft'' operators to the future. Although this sounds
like a cheat, from the viewpoint of phenomenology at the electroweak
scale it certainly isn't: the effective theory allows for the most 
general possible case, and hence provides the ideal tool for testing
the idea of weak scale supersymmetry, without needing to know anything
about the origin of supersymmetry breaking at shorter distances. 

Supersymmetry breaking can trigger
electroweak symmetry breaking as a heavy top quark effect: the large
top quark Yukawa coupling provides a controlled, negative radiative 
correction to the mass squared parameter for the Higgs boson $h_2$. 
The theory as it stands
has a physical Higgs boson mass that is lighter than the mass of the
$Z$ boson in tree approximation. This is not necessarily a problem,
since radiative corrections to the Higgs mass from the top squark may
be large.

The basic supersymmetrization described above, with the superpotential
of (\ref{eq:W}) cannot be the
whole story, since it has three very clear conflicts with data.
\begin{itemize}
\item The only unknown parameters of the Higgs potential are the two
  soft mass squared parameters, $m_1^2, m_2^2$. There are no values
  for these parameters that lead to stable, non-zero vacuum expectation
  values (vevs) for both Higgs doublets, as is  required to give
  masses to all the quarks and charged leptons. 
\end{itemize}
Even if this problem is solved, the theory possesses
two particles which are clearly experimentally excluded.
\begin{itemize}
\item  The interactions of (\ref{eq:W}) are invariant under a global 
  Peccei--Quinn symmetry \cite{PQ}, that is spontaneously broken by the
  vevs of $h_{1,2}$ leading to an electroweak axion
  \cite{Weinberg,Wilczek}. This axion is excluded, for example by K meson decays 
  and astrophysics.
\item The theory possesses two integrally charged Dirac fermions: the
  charginos. In the limit that the two Higgs vevs are equal and the
  supersymmetry breaking gaugino mass terms are ignored these two
  fermions are degenerate and have a mass $M_W$. Allowing the vevs to
  differ and introducing gaugino mass terms removes the degeneracy,
  so that one of the charginos becomes heavier than $M_W$ and the
  other lighter. Radiative corrections are small, and this
  light chargino is excluded by the LEP experiments.
\end{itemize}

Is it possible to be agnostic about how these two particles get heavy? 
Can we study an effective field theory
where we simply add by hand an axion mass and a mass for the light chargino?
The situation would then be very similar to supersymmetry breaking,
and we could postpone worrying about the origin of such masses. For
the chargino the answer is no: such a mass term breaks supersymmetry
in a way that damages the radiative structure of the theory. One must
address the origin of the light chargino mass via fully supersymmetric 
interactions at the weak scale. However, for the axion mass the answer
is yes: experiment requires only a small axion mass, and a small axion
mass can be added without doing violence to the theory. It is this
observation that leads us to a new framework for weak scale
supersymmetry. Explicit breaking of the  Peccei--Quinn symmetry 
may originate from scales far above the electroweak 
scale, even at the Planck scale, or it may occur very weakly at
the TeV scale. Either way, we need not address this physics
to pursue the phenomenology of supersymmetry at the electroweak scale.
A small soft breaking of Peccei--Quinn symmetry is analogous to soft
breaking of supersymmetry.
As far as the weak scale effective theory is concerned, the only
consequence is the appearance of an axion mass. 

In section II we present the theoretical framework that incorporates this
idea, and compare it to standard supersymmetric electroweak theories. In
section III the Higgs potential is studied and a vacuum is found where 
a $\mu$ parameter of order the supersymmetry breaking scale is 
geneerated--a result obtained earlier by Ciafaloni and Pomarol \cite{CP}. 
The scalar and fermion spectrum of the Higgs sector is also
discussed. Limits on our theory from LEP experiments, other
terrestrial experiments and from astrophysical observations are
studied in sections IV, V and VI, respectively. Cosmological signals
from BBN and CMB eras are discussed in section VII, as well as LSP
dark matter. In section VIII we discuss signatures at future colliders
and draw conclusions in section IX.

\section{Theoretical Framework}
\label{sec:theory}

The basic supersymmetrization discussed above is very economical. The
supersymmetric gauge interactions and the supersymmetric Yukawa
interactions of (\ref{eq:W}) involve the same number of parameters as
their non-supersymmetric counterparts in the standard
model. Supersymmetry breaking is described in a phenomenological way
by adding the most general set of soft operators: gaugino masses,
scalar masses and a trilinear scalar interactions for each term in 
(\ref{eq:W}). These $A$ terms, and the gaugino masses, imply that the
theory has no $R$ symmetry, but there are other global
symmetries. There are 7 chiral multiplets of differing form and
(\ref{eq:W}) possesses 3 interactions, so the theory possesses 
4 flavour symmetric $U(1)$ symmetries: 
gauged hypercharge, together with the global baryon number, $B$, 
lepton number, $L$, and Peccei--Quinn, $PQ$, symmetries. 
Indeed, given the field content, the basic supersymmetrization 
is the most general softly broken supersymmetric theory 
with these symmetries.

The MSSM provides an economical solution to the three problems,
discussed in the introduction, of the
basic supersymmetrization. No new fields
are added, but the most general set of $PQ$ breaking interactions are
added
\begin{equation}
\Delta W_{MSSM} = \mu e^{-i \phi_B} \; H_1 H_2
\label{eq:WMSSM}
\end{equation}
together with the soft supersymmetry breaking interaction
\begin{equation}
\Delta V_{MSSM} = - (\mu B \;  h_1 h_2 + h.c.) .
\label{eq:VMSSM}
\end{equation}
The parameters $\mu$ and $B$ are real, and we have chosen to write the 
physical phase in the superpotential so that the scalar potential is real.
All three problems are solved: the soft mass term allows a stable
vacuum with both vevs non-zero, the chargino mass is proportional
to $\mu$ and the axion mass is proportional to $B$. However, the
parameter $\mu$ introduces its own problem. The whole idea of having
supersymmetry at the weak scale is to trigger electroweak symmetry
breaking from supersymmetry breaking. But $\mu$ gives the Higgs bosons
a supersymmetric mass. Since $\mu$ is allowed by the symmetries of the
theory, what stops it from being huge? Why
should it have anything to do with the mass parameters appearing in
the supersymmetry breaking interactions? In certain theories it is
possible to understand that $\mu$ is itself triggered by supersymmetry
breaking, and the fact that it happens to be supersymmetric is
essentially accidental \cite{GM}. However, this applies to a
restricted set of theories of supersymmetry breaking, and in general
the mystery of why $\mu$ is comparable in size to the soft parameters
is a failing of the MSSM.

The obvious solution to the $\mu$ problem is to promote $\mu$ to a
chiral superfield, $S$, which is a singlet under the known gauge
interactions. The desired mass parameter then results when
supersymmetry breaking triggers $S$ to have a vev of order the
supersymmetry breaking scale. The immediate problem is that this
reintroduces an electroweak axion. To give the axion a mass a second
supersymmetric interaction is needed, so that the next-to-minimal
model (NMSSM) is described by the superpotential 
\begin{equation}
\Delta W_{NMSSM} = \lambda e^{-i \phi_{A_\lambda}} \; S H_1 H_2 + \kappa
 e^{-i \phi_{A_\kappa}} \; S^3
\label{eq:WNMSSM}
\end{equation}
and the soft operators
\begin{equation}
\Delta V_{NMSSM} = m_S^2 \; s^*s -  (A_\lambda  \lambda \; s h_1 h_2 
           +  A_\kappa \kappa \; s^3+ h.c.),
\label{eq:VNMSSM}
\end{equation}
which together contain 7 real parameters.
This theory is also completely realistic: there is a stable
electroweak symmetry breaking vacuum with the chargino mass deriving
from the $S H_1 H_2$ interaction and the axion mass from the $S^3$
interaction. However, the parameter space of this theory is
significantly larger than that of the MSSM --- it is not even possible
to remove phases from the Higgs potential. 

We construct a theory where the $\mu$ parameter is again promoted to a
singlet chiral superfield $S$, but introduce an alternative symmetry
structure. At the weak scale the U(1) PQ symmetry is only 
an approximate symmetry, with small explicit breaking 
in addition to that from the QCD anomaly. 
This explicit breaking could arise at some high mass
scale, $M$, much larger than the weak interaction scale $v$,
such that the renormalizable interactions below $M$, both
supersymmetric and supersymmetry breaking, possess U(1) PQ symmetry 
as an accidental symmetry of the low energy theory.
The renormalizable superpotential is then the most general
allowed by the PQ symmetry, so that the operators $H_1 H_2, S, S^2, S^3$
are all forbidden. All PQ breaking is suppressed by inverse powers of
the large mass scale $M$. Alternatively, the PQ breaking could arise
as a very small effect in the renormalizable interactions, such as
$S,S^2$ or $S^3$. Either way, we assume that at the weak scale the
explicit PQ symmetry breaking is small enough that its only relevance
to data is to give a mass to the axion.
The resulting theory is described by the superpotential
\begin{equation}
\Delta W = \lambda e^{-i \phi_{A_\lambda}} \; S H_1 H_2
\label{eq:DeltaW}
\end{equation}
together with the soft operators
\begin{equation}
\Delta V = m_S^2 \; s^*s + \frac{1}{2} m_G^2 \; GG 
   -  (A_\lambda \lambda\; s h_1 h_2 + h.c.),
\label{eq:DeltaV}
\end{equation}
where $m_G^2$ is the mass squared of the pseudo-Goldstone boson $G$, and is
taken positive and much smaller than the scale of supersymmetry
breaking, and without loss of generality $\lambda$ and $A_\lambda$ are
taken real. In the
limit that $m_G \rightarrow 0$, $G$ becomes the axion and laboratory
data and astrophysical constraints require that the axion decay
constant, and therefore the $s$ vev, be larger than $10^{10}$ GeV.
For the Higgs doublets to be at the weak scale we need $\lambda <
10^{-8}$: we
recover the MSSM with the $\mu$ problem. In our theory we take $m_G
> 300$ keV, so that the laboratory and astrophysical constraints
are avoided. This mass is large enough that the strong CP problem is
not solved, so we do not refer to $G$ as the axion. 
The superpotential (\ref{eq:DeltaW}), with $\mu$ parameter promoted 
to a dynamical field, was studied by Ciafaloni--Pomarol \cite{CP}.
They were motivated by the twin problems of the $\mu$ problem and the 
doublet--triplet splitting of grand unified theories. 

In addition to $m^2_{1,2}$, our theory has 5 parameters in the
electroweak sector (those in (\ref{eq:DeltaW}, \ref{eq:DeltaV})), 
which is intermediate between the MSSM 
(the 3 parameters of (\ref{eq:WMSSM}, \ref{eq:VMSSM})) 
and the NMSSM (the 7 parameters of (\ref{eq:WNMSSM}, \ref{eq:VNMSSM})).
However, one of our parameters, $m_G$, is small and does not 
enter into the physics of neutralinos or charginos. 
For electroweak symmetry breaking we only have 4 parameters. 
In fact, in this paper we study the case that $m_S$ is
also irrelevantly small, since this gives the desired potential
minimization, reducing our parameters to 3 --- the same as the
MSSM. Although we do not pursue it in this paper, a further reduction
in the parameter space of our theory is possible. If
we assume universality of the trilinear $A$ parameters, then the
parameters $A_\lambda$ and $\phi_{A_\lambda}$ are not new but are 
the same as parameters already introduced in the basic
supersymmetrization. This is not true of the $B$ and $\phi_B$
parameters of the MSSM, since the interaction is bilinear 
not trilinear.
In this case, our theory possesses a single new parameter for
electroweak symmetry breaking, beyond those of the basic
supersymmetrization. 

There are many models that lead to our framework at the electroweak
scale, and hence to the phenomenology discussed later. 
Although we remain agnostic about the  physics that leads to $m_G$ 
in later sections, it may be useful to give explicit examples here. 
The continuous U(1) PQ symmetry may be absent at high energies and 
appear as an accidental symmetry of the renormalizable interactions, 
just as in the case of lepton number symmetry. 
Consider the most general superpotential under a $Z_4$ symmetry and 
$Z_6$ $R$ symmetry, with superfield charges shown 
in Table~\ref{tab:PQRcharge}.  
\begin{table}[t]
 \begin{center}
\caption{\label{tab:PQRcharge}The charge assignment of the 
discrete Peccei--Quinn symmetry and $R$ symmetry. } 
 \begin{tabular}{c|c|c|c|c}
   & $\; Q,L\; $ & $\; U,D,E\; $ & $\; H_1,H_2\; $ & $\; S\; $ \\
\hline
  PQ charge (mod 4) & $1/2$ & $0$ & $-1/2$ & $1$ \\
  R $\;$  charge (mod 6) & $1$ & $1$   & $ 0$   & $2$ \\
  \end{tabular}
 \end{center}
\end{table}
It consists of (\ref{eq:W}) and (\ref{eq:DeltaW}) at the renormalizable
level, and  $[S^4/M + LL H_2 H_2/M_N]_F$ at the next order.
The above discrete symmetries forbid baryon and lepton 
number violating operators such as renormalizable $[DUD+DQL+LEL]_F$ 
and dimension 5 $[QQQL+UUDE]_F$,\footnote{Note that similar 
discrete symmetries are also necessary in the MSSM, to avoid proton
decay.} while allowing Majorana neutrino masses.
The U(1) PQ symmetry appears because of the absence of 
$H_1H_2, S, S^2, S^3$ operators in the superpotential, and is broken 
by the dimension 5 operator $[S^4/M]_F$ and a soft supersymmetry 
breaking interaction $A_S s^4/M$, where $M$ is a large mass scale.
These explicit breaking operators give rise to the mass of 
the pseudo-Goldstone boson.
\begin{equation}
m_G^2 \approx \lambda \frac{\sin (2\beta)}{2} \frac{v^2 v_s}{M} 
            + \frac{A_S v_s^2}{M}
\label{eq:mG}
\end{equation}
where $\vev{s} = v_s$, $\tan \beta = v_2/v_1$ and 
$v = \sqrt{v_1^2 + v_2^2} = 175$ GeV.
We will later find that $v_s$ is significantly larger than $v$,
so that $m_G$ originates dominantly from the supersymmetry breaking 
operator $A_S S^4/M$:
\beq
m_G \approx 10 \; \mbox{keV} \frac{v_s}{\mbox{TeV}} 
  \sqrt{\frac{A_S}{100 \mbox{GeV}}}
  \sqrt{\frac{10^{18} \mbox{GeV}}{M}}.
\label{eq:mGlim}
\eeq
Thus, $m_G$ is expected to be heavy enough to satisfy laboratory and 
astrophysical constraints, when the global PQ symmetry is 
broken at the Planck scale. 

The other class of examples that falls into our effective theory 
has a U(1) PQ symmetry at high energies that is weakly broken. 
This is just like the Froggatt--Nielsen idea, which accounts for 
the small Yukawa coupling constants, and as a consequence, 
the small pion masses.
Consider adding the operator $[\zeta S]_F$ with a dimension 2
parameter $\zeta$ much smaller than $v^2$. 
A realistic mass for the Goldstone boson can be obtained, while other
corrections to the scalar potential are negligable. 
Although we do not present an explicit high-energy model, 
$\zeta$ could be of order $m_{3/2}^2 M/M_{\rm pl}$, 
where $m_{3/2}$ is the gravitino mass, $M_{\rm pl}$ 
the Planck scale, and $M$ the scale where the PQ symmetry is broken. 
Clearly, $\zeta \ll v^2$ 
when $M/M_{\rm pl}\ll 1$ or in the case of gauge mediation.

The NMSSM also falls into our framework in the limit of $\kappa \ll 1$.
It is known that the U(1) PQ symmetry is restored and 
a $\mu$ parameter of order of the electroweak scale is maintained 
in the limit $\lambda \rightarrow 0$, $\kappa \rightarrow 0$, 
$\lambda/\kappa \rightarrow$ finite, and 
$m_S^2 \approx v^2$, but this is not the limit 
of interest to us. The operator $\kappa [S^3]_F$ 
plays an important role not only in yielding the mass of the Goldstone
boson, but also in the stabilization of the scalar potential of the 
Higgs sector in this limit.\footnote{In this limit a fine-tuning 
is required to obtain a stable vacuum.} Instead, we consider 
the limit $\kappa/\lambda \ll 1$, where  $\kappa [S^3]_F$ 
does not contribute to the Higgs sector potential but only to 
the mass of the Goldstone boson. A small value for $\kappa$ 
might originate from $M/M_{\rm pl}$ where 
$M$ is, for instance, a vev of a field whose PQ charge is $-3$. 
The operator $\kappa [S^3]_F$ is no longer responsible for the 
stabilization of the vacuum, as opposed to the NMSSM.
In the next section, we will see how in our limit the vacuum is 
stabilized and $\mu$ remains of order of the electroweak scale. 
We stress that our effective theory approach, 
adding just the mass of the Goldstone boson by hand, 
is valid for all the examples discussed above.

\section{The Higgs Sector}
\label{sec:higgssector}

\subsection{The Higgs Potential}

In our theory the spontaneous breaking of electroweak and PQ
symmetries is governed by the potential
\begin{eqnarray}
V(s,h_1,h_2) & = & m_S^2 \; s^\dagger s 
                 + m_1^2 \; h_1^\dagger h_1 
                 + m_2^2 \; h_2^\dagger h_2 
                 - (A \lambda \; s h_1 h_2 + h.c.) \nonumber \\
    & & + \frac{\bar{g}^2}{2} \; (h_1^\dagger h_1 - h_2^\dagger h_2)^2 
        + \lambda^2 \; s^\dagger s (h_1^\dagger h_1 + h_2^\dagger h_2)
        + \lambda^2 \; h_1^\dagger h_1 h_2^\dagger h_2,
\label{eq:V}
\end{eqnarray}
where $\bar{g}^2 = (g'^2 + g^2)/4$, and $A$ (previously called
$A_\lambda$) is real.
The minimization equations for the vacuum $\vev{h_1} = (v_1,0),
\vev{h_2} = (0,v_2), \vev{s} = v_s$ are given by
\beqa
\left[ m_S^2 + \lambda^2 (v_1^*v_1 + v_2^* v_2) \right] v_s &=&
A \lambda v_1^* v_2^*  \label{eq:min-s}\\
\left[ m_1^2 + \lambda^2 (v_s^*v_s + v_2^* v_2) + 
\bar{g}^2 (v_1^*v_1 - v_2^* v_2) \right] v_1 &=&
A \lambda v_s^* v_2^* \label{eq:min-h1}\\
\left[ m_2^2 + \lambda^2 (v_s^*v_s + v_1^* v_1) -
\bar{g}^2 (v_1^*v_1 - v_2^* v_2) \right] v_2 &=&
A \lambda v_s^* v_1^*.
\label{eq:min-h2} 
\eeqa
Since all quantities in the square parentheses are real, 
$v_1v_2v_s$ is real. We may choose
$v_1$ real by an electroweak gauge transformation.
The phase of $v_2$ (or that of $v_s^*$) is not determined. 
This vacuum degeneracy is due to the spontaneous breaking 
of the global PQ symmetry.
The explicit breaking of this symmetry lifts the degeneracy, and 
determines the phase. If the breaking is 
due to the soft supersymmetry breaking operator $A_S s^4 /M$, 
$v_2$ and $v_s$ are real, when the phase of $S$ is chosen  
so that $A_S$ is real.
We assume that all $v_{1,2,s}$ are real in the following.

A physically acceptable vacuum must have all three vevs non-zero, in
which case these equations can be solved for $\tan \beta = v_2/v_1$,
the ratio of the electroweak breaking vevs, 
\beq
\sin 2 \beta = \sqrt{2} \sqrt{(1 + \xi) - (1 + \xi)^2 \left(
  \frac{m_1^2 + m_2^2}{A^2} + \frac{\lambda^2 v^2}{A^2} \right)},
\label{eq:tanbeta}
\eeq
where $\xi \equiv m_S^2/\lambda^2 v^2$, the vev of $s$
\beq
v_s = \frac{A}{2 \lambda} \sin (2 \beta) \frac{1}{1 + \xi}
\label{eq:vs}
\eeq
and the $Z$ boson mass
\beq
M_Z^2 = - \frac{m_1^2 - m_2^2}{ \cos (2 \beta)} + \lambda^2 v^2 -
A^2.
\label{eq:mz}
\eeq

\subsection{A Large Singlet vev}

Suppose that the singlet vev is small: $v_s \ll v$. In this case
we have a two Higgs doublet theory where the coupling of $G$ to the up
quark sector is proportional to $\cot \beta$ and the coupling of $G$
to the down sector is proportional to $\tan \beta$. If $G$ is
sufficiently light it will be produced in the decays of $K, \Psi$ and
$\Upsilon$ mesons and the theory will be excluded. Hence we must
either give $G$ a mass in the GeV range, or we must take $v_s$ somewhat
larger than $v$. In this paper we choose to focus on the latter
case, since it will allow us to study the maximum possible range of
$m_G$. For $m_G \simlt 50$ MeV, the constraint from $K^+$ decays
requires $\sqrt{2} v_s \simgt 50$ TeV, or $v_s/v \simgt 200$.

From (\ref{eq:vs}) we see that there are three possible ways 
of obtaining $v_s \gg v$. 
We may fine tune $m_S^2$ close to $-\lambda^2 v^2$, so that
$\xi$ is close to $-1$; a possibility that we ignore. 
We may take $A$ large compared with $v$; but from (\ref{eq:mz}) 
we see that fine tuning is then required to keep the $Z$ boson light. 
Hence we study the final option of small $\lambda$. 
In fact, small $\lambda$ by itself is not sufficient: 
if $m_S^2$ is of order $v^2$, then $\xi \sim 1/\lambda^2$ 
becomes large and $v_s \sim \lambda A$ becomes small. 
We must study the limit
\beq
 \lambda \ll 1, \quad {\rm and} \quad |m_S^2| \lsim \lambda^2 v^2 
\quad i.e. \; \;  |\xi | \lsim {\cal O}(1).
\label{eq:limit}
\eeq
We note that this small $|m_S^2|$ is at least technically natural.
Radiative corrections to $m_S^2$ are of order 
$(\lambda^2/16\pi^2)v^2$, and even $|m_S^2| < \lambda^2 v^2$, 
i.e., $|\xi | < 1$, is also technically natural.
An initial condition with small $|m_S^2|$ can be set by 
e.g., gauge mediation \cite{CP}, or possibly by gaugino mediation, 
and hence we consider that this is quite a plausible assumption.

In this limit we can drop the $\lambda^2 v^2$ terms from
(\ref{eq:tanbeta}) and  (\ref{eq:mz}), and set $\xi$ to zero. 
It follows that 
\beqa
\sin 2 \beta & \simeq & \sqrt{2} \sqrt{1  - 
  \frac{m_1^2 + m_2^2}{A^2}} \label{eq:approxbeta} \\
v_s & \simeq & \frac{A}{2 \lambda} \sin 2 \beta \label{eq:approxvs} \\
M_Z^2 & \simeq & - \frac{m_1^2 - m_2^2}{\cos 2 \beta} - A^2.
\label{eq:approxmz}
\eeqa
We note again that large $v_s$ is a direct consequence of 
a rather small $\lambda$. Ignoring phases,
the electroweak sector of the MSSM is controlled by four parameters
$(m_1^2, m_2^2, \mu,B)$. Our theory similarly has four parameters
controlling the electroweak sector $(m_1^2, m_2^2, \lambda,A)$, which
can be translated into $(M_Z, \tan \beta, v_s,A)$ by the minimization
constraints. 

The conditions on the parameters of the theory for successful
electroweak symmetry breaking are the following\footnote{We 
thank B.~Feldstein for a useful discussion.}:
from (\ref{eq:approxbeta}) and (\ref{eq:approxmz}), 
\begin{eqnarray}
 & & \frac{A^2}{2} < m_1^2 + m_2^2 < A^2,  \label{eq:ewsbcondA} \\
 & & (m_1^2 - m_2^2)^2 > \cos^2 (2\beta) A^4. \label{eq:ewsbcondB}
\end{eqnarray}
Given the vev of $s$ (\ref{eq:approxvs}), the minimization equations 
(\ref{eq:min-h1}, \ref{eq:min-h2}) do not have a solution 
if the inequalities (\ref{eq:ewsbcondA}) are not satisfied.
The two inequalities correspond to 
$\tan \beta = 1$ (D-flat direction) and $\tan \beta = \infty$.
Unlike in the MSSM, large $\tan \beta$ is a fine tune in our 
theory: a value of $\tan \beta = 30$ requires that $A^2$ be fined tuned
equal to $(m_1^2 + m_2^2)/2$ to within one part in a thousand. 
We will be interested in moderate values of $\tan \beta$.
If the condition (\ref{eq:ewsbcondB}) were not satisfied, 
$M_Z^2$ would be negative.

\subsection{An Effective $\mu$ Parameter}

Having chosen parameters to ensure that $h_1, h_2, s$ all acquire
vevs, it is frequently more convenient to exchange the parameter 
$\lambda$ for the derived quantity $\mu \equiv \lambda v_s$
\beq
\mu = \frac{A}{2} \sin (2 \beta).
\label{eq:mu}
\eeq
One reason for doing this is that large $v_s$ is often 
accompanied by small $\lambda$ in various expressions, and 
the combination $\mu = \lambda v_s$ is almost like the 
$\mu$-parameter of the MSSM.
The effective $\mu$-parameter is moderate in this theory, 
since $v_s \propto \lambda^{-1}$.
The mass of the lightest chargino is directly related to $\mu$, 
and there is a direct experimental limit $\mu \gsim 120$ GeV.
Other features of the theory are also apparent from (\ref{eq:mu}).
For example, for large $\tan \beta$, $A = \mu \tan \beta$ also 
becomes large, so that there is also a fine tune 
in the relation (\ref{eq:approxmz}) for $M_Z$.

Let us replace $s$ with its vev $v_s$ in the scalar potential 
(\ref{eq:V}) to obtain a potential only for $h_1$ and $h_2$ 
with a fixed value of $s$. 
For the parameter ranges of interest to us, we can ignore 
the $ \lambda^2 \; h_1^\dagger h_1 h_2^\dagger h_2$ operator 
in the potential, so that (\ref{eq:V}) can be rewritten in
the form
\beq
V(\vev{s},h_1,h_2)  \simeq   
   (m_1^2 + \mu^2)  h_1^\dagger h_1 + (m_2^2 + \mu^2) h_2^\dagger h_2 
 - (A \mu \;  h_1 h_2 + h.c.) 
 + \frac{\bar{g}^2}{2} \; (h_1^\dagger h_1 - h_2^\dagger h_2)^2.
\label{eq:Vapprox}
\eeq
This has precisely the form of the potential in the MSSM, with 
\beq
B=A. 
\label{eq:B}
\eeq
The familiar minimization equations of (\ref{eq:Vapprox}) 
\beq
\sin (2 \beta) = \frac{2 \mu B}{m_1^2 + m_2^2 + 2 \mu^2}
\label{eq:tanbetaapprox}
\eeq
and
\beq
\frac{M_Z^2}{2} = \frac{m_1^2 - m_2^2 \tan^2 \beta}{\tan^2 \beta - 1}
- \mu^2
\label{mzapprox}
\eeq
are identical to (\ref{eq:approxbeta}) and (\ref{eq:approxmz}), 
under (\ref{eq:mu}) and (\ref{eq:B}), as it should be.
The conditions for the successful electroweak symmetry breaking 
\begin{eqnarray}
& & 2 B\mu < m_1^2 + m_2^2 + 2 \mu^2, \\
& & (m_1^2 + \mu^2)(m_2^2 + \mu^2) < (B \mu)^2
\end{eqnarray}
of the MSSM are identical to the first inequality of 
(\ref{eq:ewsbcondA}) and (\ref{eq:ewsbcondB}), respectively, 
under (\ref{eq:mu}) and (\ref{eq:B}).

The $\mu$-parameter of the MSSM is given by $\lambda v_s$, as 
in the NMSSM.
However, the crucial difference from the NMSSM is that 
$\mu$ of order of the electroweak scale is guaranteed 
independently of the value of $\lambda$. 
The plausible assumption $|m_S^2| < \lambda^2 v^2$ ensures 
that $v_s \propto \lambda^{-1}$, and that 
$\mu$ is given by (\ref{eq:mu}) independent of $\lambda$.
Thus, the $\mu$-problem is solved in this theory in a very different
way from the solution of the NMSSM \cite{CP}.
We also see shortly that small $|m_S^2|$ also ensures 
vacuum stability.

\subsection{The Scalar Spectrum and Mixing}

The basic properties of the scalars in the Higgs sector 
are below; most of these results can be obtained from studies of the
NMSSM by taking the limit of vanishing $[S^3]_F$ coupling 
\cite{Zerwas,MN}.

There are seven on-shell scalar particles in the  Higgs sector.
Two of them form an electrically charged scalar. 
Of the five neutral scalars,
three are CP-even and two CP-odd.

\noindent {\bf Charged scalar} \\
The charged scalar comes from the two Higgs doublets just as 
in the MSSM. Its mass eigenvalue is given by 
\begin{equation}
M_{H^{\pm}}^2 = M_A^2 + M_W^2 - \epsilon^2 \mu^2, 
\end{equation}
where $\epsilon \equiv v/v_s \ll 1$ and 
\begin{equation}
M_A^2 \equiv \frac{2}{\sin (2 \beta)} \mu A = \frac{1}{1+\xi} A^2.
\label{eq:A-mass}
\end{equation}

\noindent {\bf Neutral CP-even scalars} \\
The three CP-even neutral scalars come from the real-scalar parts of 
$h_1,h_2$ and $s$. 
Taking a basis $\bar{s} = (s_1,s_2,s_3)$ determined by 
\begin{eqnarray}
\Re \; h_1 & = & \frac{1}{\sqrt{2}}(- s_1 \sin \beta + s_2 \cos \beta) 
                + v_1, \\
\Re \; h_2 & = & \frac{1}{\sqrt{2}}(s_1 \cos \beta + s_2 \sin \beta)
                + v_2, \\
\Re \; s \; \; & = & \frac{1}{\sqrt{2}} s_3 + v_s,
\end{eqnarray}
the mass matrix is given by 
\begin{eqnarray}
M^2_+ & = & \left(  \begin{array}{ccc}
  M_A^2 +M_Z^2 \sin^2(2\beta) & - M_Z^2 \sin (2\beta) \cos (2\beta) 
       & 0 \\
  - M_Z^2 \sin (2\beta) \cos (2\beta) & M_Z^2 \cos^2 (2\beta) & 0 \\
	0  & 0 & 0 \\
	\end{array}\right) \nonumber \\ & + &
  \epsilon
         \left(  \begin{array}{ccc}
	     0 & 0 & - \cos (2\beta) \\ 
	     0 & 0 & \sin (2 \beta) 
               \left(\left(\frac{1}{1+\xi}\right)^2 - 1 \right) \\
	     - \cos (2\beta) & \sin (2 \beta) 
               \left(\left(\frac{1}{1+\xi}\right)^2 - 1 \right) & 0 \\
		   \end{array}\right)  M_A^2 \frac{\sin (2 \beta)}{2}
        \nonumber \\  
&+& \epsilon^2 \left( \begin{array}{ccc}
    - \sin^2 (2\beta) \mu^2 & \sin (2\beta) \cos (2\beta) \mu^2 & 0 \\ 
    \sin (2\beta) \cos (2\beta) \mu^2 & \sin^2 (2\beta) \mu^2 & 0 \\ 
	      0 & 0 & M_A^2 \left(\frac{\sin (2\beta)}{2}\right)^2 \\
		   \end{array}\right). 
\label{eq:scalmass}
\end{eqnarray}
Since we are interested in the parameter region\footnote{
The parameter region studied extensively in the appendix of 
\cite{Zerwas} has $M_A^2 \gg M_Z^2$ and $\tan \beta \gg 1$. We do not 
assume either of these conditions, and hence the region we are 
interested in is completely different from the one in \cite{Zerwas}.
Reference \cite{MN} is interested in 
$\cos \beta_s \equiv \epsilon \equiv v/v_s \ll 1$,
as we are, and contains an expansion similar to (\ref{eq:scalmass}).} 
with $\lambda \ll 1$, or 
equivalently $\epsilon \equiv v/v_s = \lambda v/\mu \ll 1$, 
the first term of (\ref{eq:scalmass}) dominates, 
with small corrections of order $\epsilon$ and $\epsilon^2$. 
The first term of (\ref{eq:scalmass}) is the scalar mass matrix of the 
MSSM, with $B=A$.

The mass-squared eigenvalues are determined as follows. 
The two larger eigenvalues are almost those of the MSSM, 
with corrections of ${\cal O}(\epsilon^2,\xi)$, and the corresponding 
mass eigenstates are very much like the neutral CP-even 
Higgs scalars $H$ and $h$ of the MSSM. 
The mass eigenvalue of the $h$-like state is not larger than 
$m_Z |\cos (2\beta)|$ at tree level.  
The smallest mass-squared eigenvalue turns out to be 
$(\lambda v \sin (2 \beta))^2 
(1+ {\cal O}(\epsilon^2,\xi))$, and is positive as long as 
$|\xi | \lsim 1$, i.e., $|m_S^2| \lsim \lambda^2 v^2$. 
Thus, the vacuum instability discussed in \cite{Zerwas} is avoided 
if $|m_S^2|$ is sufficiently small.
We stress that a stable vacuum does not require a fine tuning 
in this theory.
The mass eigenvalue $\sin (2\beta) \lambda v$ is quite small, and 
remains well below $M_Z$ 
even after radiative corrections are taken into account, 
since the corrections involve the small coupling constant $\lambda^2$.

Denoting the mass eigenstates by $\bar{H} = (H_3,H_2,H_1)$, 
$H_3$ and $H_2$ are almost $H$ and $h$ of the MSSM, and are therefore
expected to have very similar properties to $H$ and $h$. The very
light scalar $H_1$ is a new feature of this theory and is 
almost contained in $\Re \; s$.
Defining the orthogonal transformation between the two bases by
$\bar{H} = O \bar{s}$, 
we find that the $H_1$ components in the doublet states $s_{1,2}$ are
given by
\begin{eqnarray}
  O_{H_1 s_1} & = & \frac{\sin (4\beta)}{4}\epsilon 
          + {\cal O}(\epsilon^3,\epsilon \xi), \\
  O_{H_1 s_2} & = & \frac{\sin^2 (2\beta)}{2} \epsilon 
          + {\cal O}(\epsilon^3, \epsilon \xi),   
\label{scalarmixings}
\end{eqnarray}
All the interactions of $H_1$ are of order $\epsilon$ or smaller, 
either via these mixings or via the coupling $\lambda$.

\noindent {\bf Neutral CP-odd scalars} \\
The two CP-odd scalars come from the three phase directions of 
the three complex scalars $h_1,h_2$ and $s$. 
One of them, $A$, is massive because of the potential 
(\ref{eq:DeltaV}).
Its mass eigenvalue is given by $M_A^2$ defined 
in Eq.~(\ref{eq:A-mass}), which is the same result 
as that of the MSSM with $B=A$.
The other state is the pseudo-Goldstone boson, $G$.

The mass eigenstates are described as follows.
Let us first take a basis $(p_1,p_2)$:
\begin{eqnarray}
  p_1 \; :  & \quad & \delta (h_1,h_2,s) 
     = \left(v_1 e^{i \tan \beta \frac{\varphi_1}{\sqrt{2}v}}, 
             v_2 e^{i \cot \beta \frac{\varphi_1}{\sqrt{2}v}},
             0 \right), \\
  p_2 \; :  & \quad & \delta (h_1,h_2,s) 
     = \left(0,0,v_s e^{i \frac{\varphi_2}{\sqrt{2}v_s}}\right).
\end{eqnarray}
Then the massive state, $A$, corresponds to the degree of freedom
\begin{equation}
 (\varphi_1,\varphi_2) \propto 
     \left(\frac{v}{v_1 v_2},\frac{1}{v_s} \right), 
\label{eq:A-direct}
\end{equation}
while the Goldstone boson corresponds to 
\begin{equation}
  \frac{1}{\sqrt{2}}(\varphi_1,\varphi_2) 
      = \left(-\frac{v_1 v_2}{v},v_s \right)\frac{G}{F_G},
\label{eq:G-direct}
\end{equation}
where 
\begin{equation}
 F_G \equiv \sqrt{2}
           \sqrt{\left(\frac{\sin (2 \beta)}{2} \right)^2 v^2 + v_s^2} 
  \approx \sqrt{2} v_s
\end{equation}
is the decay constant of the Goldstone boson.\footnote{
\label{fn:normalization} Note that 
we have adopted a normalization of the decay constant different 
from the one common in the literature of 
the electroweak (Peccei--Quinn--Weinberg--Wilczek) axion, 
where $F_G = \sqrt{2} v = 246$ GeV. 
The difference is $\sin( 2 \beta)/2$.} 
The mixing angle $\theta_-$ of the orthogonal rotation 
between the $(p_1,p_2)$ basis and the mass eigenstate basis is 
given by 
\begin{equation}
 \tan \theta_- =  \epsilon \frac{\sin (2\beta)}{2}.
\end{equation}
The mixing angle is small when $\epsilon \ll 1$. 
The Goldstone boson is contained mainly in $p_2$, the phase of 
the complex scalar $s$, and the massive pseudo-scalar is mainly
in $p_1$. 

The direction determined 
by (\ref{eq:G-direct}) corresponds to 
the Goldstone boson of a U(1) symmetry whose charge assignment is 
$- \sin^2 \beta, -\cos^2 \beta$ and $+1$ for $h_1,h_2$ and $s$, 
respectively.\footnote{The direction of the massive 
pseudo-scalar, i.e., (\ref{eq:A-direct}) is orthogonal 
both to the would-be Goldstone direction of the $Z$ boson and 
to the Peccei--Quinn transformation of the vacuum. 
The direction of the Goldstone boson should be a symmetry direction, 
and hence is given by a linear combination of 
the two symmetry transformations above. 
It is orthogonal both to the massive-state direction and 
to the would-be Goldstone direction of the $Z$ boson. }
This U(1) symmetry is a combination of 
the ordinary Peccei--Quinn symmetry whose charge assignment is 
$-1/2,-1/2,+1$, respectively, and a symmetry corresponding 
to the Z boson, whose assignment is $-1/2,+1/2,0$, respectively.

\subsection{The Fermion Spectrum}

The mass matrix for the charginos and the $4 \times 4$ mass matrix 
for the neutral Higgsinos and gauginos is exactly the same as in 
the MSSM, with $\mu$ parameter given by $\lambda v_s$. 
However, there is a fifth neutralino arising from the
fermion in the $S$ superfield, $\tilde{s}$. 
This s-ino mixes with the neutral
Higgsinos via the mass terms $\lambda v_1 \tilde{s} \tilde{h}_2 + 
\lambda v_2 \tilde{s} \tilde{h}_1$. 
Since these masses are a factor $\lambda = \epsilon (\mu/v)$ smaller 
than the masses of the standard neutralinos, and since 
there is no $\tilde{s}\tilde{s}$ mass term in this theory,\footnote{
What follows is also valid when the PQ symmetry is broken 
by $\kappa [S^3]_F$ and its A term, if $\kappa \lsim \lambda^3$.} 
the lightest state $\tilde{\chi}^0_0$ obtains a seesaw mass of order 
$\lambda^2 v^2 /m_{\rm SUSY}$, where $m_{\rm SUSY}$ is $\mu$ or 
gaugino masses. 
The mixing angle between the s-ino and the
doublet Higgsinos is of order $\epsilon$. 
The lightest neutralino $\tilde{\chi}^0_0$ is the lightest SUSY 
particle (LSP), and is almost identical\footnote{Thus, $\tilde{s}$ 
sometimes stands for the LSP in the following.} to $\tilde{s}$. 
From this one can obtain the order of magnitude of the s-ino (the LSP)
interactions, for example, 
the $Z \tilde{h} \tilde{s}$ coupling is of order $\epsilon$, 
the $Z \tilde{s} \tilde{s}$ coupling of order $\epsilon^2$, and 
the $G \tilde{s} \tilde{s}$ coupling is of order $\epsilon^3$.

\noindent {\bf Spectrum Summary} \\
Compared to the MSSM, this theory contains two additional light Higgs 
scalars ($H_1,G$) and a light neutralino ($\tilde{s}$). 
As the coupling constant $\lambda$ is taken small, the light states 
lie dominantly in the singlet superfield $S$, and they decouple from 
the rest of the theory. The scalar $H_1$ has a mass of order
$(400 \lambda) \times 0.3$ GeV, 
$\tilde{s}$ has a mass of order $(400 \lambda)^2 \times 0.5$ MeV, 
and is the LSP, while $m_G$ is a free parameter. 
All three states have couplings of order $\epsilon \sim \lambda$ 
or less.
The mass spectrum and interactions of the other Higgs scalars,
charginos and neutralinos closely
resemble those of the MSSM, the deviations being of order $\epsilon$.
In the MSSM these masses and couplings depend on the gaugino mass
parameters and on $\tan \beta, \mu$ and $M_A$. The same is true in our
theory, except that now $\mu$ is not an additional free parameter but
is predicted by
\beq
\mu = \frac{M_A}{2} \sin 2 \beta ( 1 +  {\cal O}(\epsilon, \xi)), 
\label{eq:mupred}
\eeq
first obtained by Ciafaloni and Pomarol \cite{CP}.
This prediction, which is an important test of this theory, arises
because the effective $\mu$ and $B$ parameters are not independent
but are related by $\mu = (B/2) \sin (2 \beta)$.

\section{Limits from LEP}
\label{sec:LEP}

The five neutral scalars couple to the $Z$ gauge boson 
through 
\begin{equation}
 {\cal L} = \frac{\sqrt{g^2+g^{'2}}}{2}M_Z Z_\mu Z^\mu s_2 
          + \frac{\sqrt{g^2+g^{'2}}}{2}Z_\mu 
               \left[s_1 \partial^\mu p_1 - p_1 \partial^\mu s_1 
               \right].
\label{eq:Zcoup}
\end{equation}
The CP-even scalar $s_3$ and the CP-odd scalar $p_2$ do not 
couple to the electroweak gauge bosons because they come 
from $s$, which is neutral under the standard-model gauge group.
The first interaction allows for $H_i$ to be singly produced via 
a virtual Z boson. This leads to a bound on the mass of $H_2$ identical
to that on the mass of $h$ of the MSSM. The cross section for $H_1$
production via $e^+ e^- \rightarrow Z H_1$ receives a suppression
factor of $\epsilon^2 / \tan^4 \beta$ from the mixing angle
$O_{H_1 s_2}$ of (\ref{scalarmixings}). (Here and below we approximate
$\sin 2 \beta/2$ as $1/\tan \beta$, which is a reasonable approximation
even for moderate $\tan \beta$.)

In the MSSM all Higgs scalars, pseudo-scalars and their superpartners
are too heavy to be produced in $Z$ decay. In our theory there is the
possibility that $H_1,G$ and $\tilde{\chi}^0_0$ are produced 
in $Z$ decay.
In practice the relevant decay modes are highly suppressed. The
amplitude for $Z \rightarrow H_1 G$ is suppressed by the mixing angles 
$O_{H_1 s_1} \theta_-$, giving a branching ratio suppressed by
$\epsilon^4/ \tan^4 \beta$, while the branching ratio for $Z
\rightarrow \tilde{\chi}^0_0 \tilde{\chi}^0_0$ is also of order 
$\epsilon^4$. 
Thus LEP data is only able to constrain $\epsilon$ to be less than 
of order $0.1$, while below we find that other limits 
are more powerful by some two orders of magnitude. 
Clearly there will be no signals for our theory in $Z$ decay.

\section{Other Terrestrial Limits}
\label{sec:terr}

The light states coming from $S$ are the prominent feature 
of this theory. 
We have seen in the previous section that they are not excluded 
by the LEP experiments.
However, processes with lower energy can also 
put constraints on the properties of such light states.
We show in section \ref{sec:terr} and section \ref{sec:astro} 
that our theory still survives other terrestrial and astrophysical 
limits, respectively. 

Among the light states, the CP-even scalar is not produced 
in processes with lower energy, because it is not much lighter 
than a GeV. 
The lightest neutralino $\tilde{\chi}^0_0$ is not produced either; 
because it is almost sterile, and moreover, it has to be 
created in pairs. 
Thus, the amplitudes creating $\tilde{\chi}^0_0$ 
are highly suppressed. 
The Goldstone boson is light enough to be created in 
various low-energy processes. Since it can be produced alone, 
such amplitudes are not suppressed very much.
Thus, we devote section \ref{sec:terr} and \ref{sec:astro} to 
the discussion of various phenomenological limits on 
the properties of this light boson.

The Goldstone boson of this theory has properties quite similar 
to those of the QCD axion, and in particular, the DFSZ-type axion 
\cite{DFSZ}. 
The major difference from the DFSZ axion is in its mass.
Although the mass of the QCD axion is given by \cite{BT} 
\begin{equation}
 m_G = m_{\pi^0}\frac{F_{\pi^0}}{F_G}N_g \frac{\sqrt{z}}{1+z}
       \simeq 18 \; \KEV \left(\frac{1 \TEV}{F_G}\right),
\label{eq:BT}
\end{equation}
where $N_g = 3$ is the number of generations, 
$z\equiv m_u/m_d = 0.56$, and $m_{\pi^0}$ and $F_{\pi^0}$ the mass 
and decay constant of $\pi^0$, respectively, 
$m_G$ is completely independent of the decay constant $F_G$ 
in our effective theory, except that it is larger than (\ref{eq:BT}).
The QCD contribution to the mass of the Goldstone boson is dominated 
over, for instance, by those from explicit breaking operators 
such as (\ref{eq:mGlim}) for $F_G$ larger than a few TeV.

Various constraints on light neutral CP-odd scalar particles 
have been discussed in the literature, and review articles 
are also available. 
But, most of the literature is motivated by the axion, and hence 
some of them are only for the PQWW axion, and  
some assume the QCD relation Eq.~(\ref{eq:BT}).
Some references are more general and obtain a conservative analysis  
by using only the coupling to photons. 
We obtain limits on our theory by re-examining the various 
constraints in the literature.
At the end of section \ref{sec:terr} and \ref{sec:astro}, 
the limits are described on the $F_G$--$m_G$ plane 
in Fig.~\ref{fig:terr} and \ref{fig:astro}, respectively.
A brief summary of the allowed region is found at the end of section 
\ref{sec:astro}.

\subsection{Low-energy effective action of the Goldstone boson}

Before discussing each limit, 
we briefly summarize the low-energy effective action 
of the Goldstone boson.
The particle contents of the effective theory well below 
the electroweak scale consist of photon, gluon, quarks, leptons, 
and the Goldstone boson $G$.
The couplings of the Goldstone boson to quarks and leptons are
given by \cite{Kim-Review, Cheng-Review}
\begin{eqnarray}
 {\cal L} & = & \frac{1}{2}\partial_\mu G \partial^\mu G + 
            \sum_j \left(\bar{u}_j i \Dsl u_j + \bar{d}_j i \Dsl d_j 
                 + \bar{e}_j i \Dsl e_j \right)
              \label{eq:leading}\\
          & & - \sum_j \left(
                m_{uj} \bar{u}_j u_j + m_{dj} \bar{d}_j d_j 
              + m_{ej} \bar{e}_j e_j  \right)     \\
          & & - \sum_j \left(
              \frac{\cos^2 \beta}{2}\frac{\partial_\mu G}{F_G}
                 \bar{u}_j \gamma^\mu \gamma_5 u_j
            + \frac{\sin^2 \beta}{2}\frac{\partial_\mu G}{F_G}
                 \left(  \bar{d}_j \gamma^\mu \gamma_5 d_j  
                       + \bar{e}_j \gamma^\mu \gamma_5 e_j 
                 \right)
                       \right) 
              \label{eq:axvct1} \\
& & + \frac{1}{2} \frac{\partial_\mu G}{F_G} \left(
       \frac{N_g}{1+z+w}\bar{u_1}\gamma^\mu \gamma_5 u_1
      +\frac{N_g z}{1+z+w}\bar{d_1} \gamma^\mu \gamma_5 d_1 
      +\frac{N_g w}{1+z+w}\bar{d_2} \gamma^\mu \gamma_5 d_2 \right), 
              \label{eq:axvct2}
\end{eqnarray}
where $u_j,d_j,e_j$ are Dirac spinors of up-type quark, 
down-type quark, and charged lepton in the $j$-th generation.
$w \equiv m_u/m_s \simeq 0.03$, and $\gamma_5$ is 
$-1 $ for left-handed spinors and $+1 $ for right-handed spinors. 
The interaction of (\ref{eq:axvct2}) comes from mass mixing with 
the $\eta^{(')}$ and $\pi^0$ mesons, where $m^2_{\pi^0}/m^2_{\eta}$ 
and $m_G^2/m^2_{\pi^0}$ have been neglected.\footnote{When $m_G^2$ 
becomes large, the coefficient of the isovector component of 
the axial current in (\ref{eq:axvct2}) is modified 
by ${\cal O}(m_G^2/m_{\pi^0}^2)$.}

The effective theory is described in terms of hadrons rather 
than quarks when the relevant energy is much lower than a GeV.
The Goldstone boson still couples to an axial vector current, 
accompanied by finite renormalization factors of order unity.
Eqs.~(\ref{eq:axvct1}) and (\ref{eq:axvct2}) are replaced by 
\begin{equation}
- \frac{\partial_\mu G}{2 F_G} \bar{\psi}\gamma^\mu \gamma_5 
   (g^{(0)}+\tau^3 g^{(1)}) \psi,
\label{eq:NNG}
\end{equation}
where $\psi=(p,n)$ is the isospin doublet of proton and neutron, and 
the coefficients $g^{(0)}$ and $g^{(1)}$ for the iso-scalar and 
iso-vector pieces are given by 
\begin{eqnarray}
g^{(0)} & = & \frac{0.166}{2}
          \left[ \left(\cos^2 \beta - \tilde{N_g} \right) + 
                 \left(\sin^2 \beta - \tilde{N_g} z \right)
          \right] 
        - 0.257  \left(\sin^2 \beta - \tilde{N_g} w \right),\\
g^{(1)} & \equiv & \frac{F_A^{(1)} = 1.25}{2}\rho^{(1)} \nonumber \\
        & = & \frac{1.25}{2} 
          \left[ \left(\cos^2 \beta - \tilde{N_g} \right) -  
                 \left(\sin^2 \beta - \tilde{N_g} z \right)
          \right]   =  \frac{1.25}{2}(\cos^2 \beta - \sin^2 \beta - 0.830),
 \label{eq:isovect-form}
\end{eqnarray}
respectively \cite{Srednicki,DBKaplan,EMOSSW}, 
where $\tilde{N_g} = N_g / (1 + z + w)$ and 
$N_g=3, z=0.56, w=0.03$ are used in the last line. 
All the above effective interactions of the Goldstone boson are 
the same as those of the non-SUSY DFSZ axion.

The anomalous coupling of the Goldstone boson with photons is given by 
\begin{equation}
{\cal L} = 
         - \left[\frac{4}{3}N_g - 1 - \frac{N_g}{3}\frac{4+z+w}{1+z+w}
           \right]
           \frac{\alpha_{QED} }{4 \pi}\frac{G}{F_G}
                       F_{\mu\nu}\tilde{F}^{\mu\nu}.
\label{eq:G-QED}
\end{equation}
The second term of the anomaly coefficient in the bracket ($-1$)
is from the charged Higgsinos. 
The numerical value of the anomaly coefficient is accidentally small 
\begin{equation}
 \left[\frac{4}{3}N_g - 1 - \frac{N_g}{3}\frac{4+z+w}{1+z+w} \right]
  = 0.113.
\label{eq:QED-anomaly}
\end{equation}
The Goldstone boson decays to two photons through the anomalous 
coupling (\ref{eq:G-QED}) with a decay rate 
\begin{equation}
 \Gamma(G \rightarrow 2\gamma) 
  = 0.113^2 \frac{ \alpha_{QED}^2}{64\pi^3}  
    \frac{m_G^3}{F_G^2}, 
\end{equation}
and a decay length
\begin{equation}
 l_{2\gamma}\gamma \equiv \Gamma^{-1}\frac{E_G}{m_G} 
  = \left(\frac{100 \; \KEV}{m_G}\right)^4 \left(\frac{E_G}{\MEV}\right)
    \left(\frac{F_G}{10 \; \TEV}\right)^2
    \times 5.6 \times 10^{14}{\rm m.} 
\label{eq:lengthGgg}
\end{equation}
If it is heavier than $2 m_e$, then it decays to $e^+ e^-$, with a
decay rate  
\begin{equation}
 \Gamma(G \rightarrow e^+ e^-) = 
   \frac{m_G}{8\pi}\frac{m_e^2}{F_G^2} \sin^4 \beta, 
\label{eq:Gtoee}
\end{equation}
and the decay length becomes 
\begin{equation}
 l_{2e} \gamma \equiv \Gamma^{-1}\frac{E_G}{m_G} = 
    \left(\frac{\MEV}{m_G}\right)^2 \left(\frac{E_G}{\MEV}\right)
    \left(\frac{F_G/(\sin^2 \beta)}{\TEV}\right)^2 
    19  {\rm m}.
\label{eq:lengthGee}
\end{equation}

The lightest neutralino $\tilde{\chi}^0_0$ is also light 
in this theory. It can be heavier or lighter than electron, 
depending on $\lambda$. 
The low-energy effective interaction between the Goldstone boson 
and $\tilde{\chi}^0_0$ will be given by 
\begin{equation}
 c \lambda^3 G \tilde{\chi}^0_0 \tilde{\chi}^0_0 + h.c., 
\end{equation}
where $c$ is a coefficient of order unity.
Since $m_{\tilde{\chi}^0_0}/F_G \approx \lambda^3$, the decay rate
$\Gamma(G \rightarrow \tilde{\chi}^0_0 \tilde{\chi}^0_0)$ is comparable to 
$\Gamma(G \rightarrow e^+ e^-)$, 
when $m_{\tilde{\chi}^0_0} \approx m_e < m_G$.
If $2 m_{\tilde{\chi}^0_0} < m_G < 2 m_e$, 
then the Goldstone boson decays dominantly to $\tilde{\chi}^0_0 \tilde{\chi}^0_0$
with a decay length 
\begin{equation}
 l_{2\tilde{\chi}^0_0}\gamma = 
     \left(\frac{\MEV}{m_G}\right)^2 
     \left(\frac{E_G}{\MEV}\right)
     \left(\frac{(1/400)^3}{\lambda^3 c}\right)^2
     \times 2 \times 10^4 {\rm m}.
\label{eq:lengthGcc}
\end{equation}

\subsection{Rare decay of mesons}

Quarkonium decays to the Goldstone boson have not been observed 
\cite{PDG}:
\begin{eqnarray}
 {\rm Br}(J/\psi \rightarrow G + \gamma) & < & 1.4 \times 10^{-5} 
    \quad (90 \% {\rm C.L.}), \\
 {\rm Br}(\Upsilon (1S) \rightarrow G +\gamma) & < & 3 \times 10^{-5}
   \quad (90 \% {\rm C.L.}).
\label{eq:Br-U-old}
\end{eqnarray}
Thus, the coupling of the Goldstone boson with charm and bottom quarks,
i.e., $ \cos^2 \beta (m_c/F_G)$ and $\sin^2 \beta (m_b/F_G)$, 
must be sufficiently small. 
The limit on the decay constant is given by \cite{Kim-Review}
\begin{eqnarray}
  F_G  & \gsim &  \cos^2 \beta \times (1 \sim 2) \; \TEV 
       \qquad ({\rm for~}m_G \ll m_c), \\
  F_G  & \gsim &  \sin^2 \beta \times 500 \; \GEV    
       \qquad \quad \; \; ({\rm for~}m_G \ll m_b). 
\end{eqnarray}

A recent experimental constraint from $K^+$ decay 
\cite{Kdecay}\footnote{The new data contains an event 
consistent with 2-body decay and sufficiently small $m_G$.} 
\begin{equation}
{\rm Br}(K^+ \rightarrow \pi^+ + G) \lsim 0.73 \times 10^{-10} 
\qquad (90 \% {\rm C.L.})
\label{eq:Kbr}
\end{equation}
provides a more stringent constraint.
The theoretical estimate of the branching ratio has 
large uncertainties, and we just quote an estimate  
from \cite{Kim-Review}\footnote{An estimate in \cite{AT} is 
roughly 30 times larger than the one quoted in the text. 
If we adopt this estimate, then $300 \; \TEV \lsim F_G$.} 
\begin{equation}
{\rm Br}(K^+ \rightarrow \pi^+ + G) \sim 3 \times 10^{-6}
  \left(\frac{250 \; \GEV}{F_G}\right)^2.
\end{equation}
Thus, we obtain a rough estimate of the lower bound 
of the decay constant:
\begin{equation}
 F_G  \gsim  50 \; \TEV.
\end{equation}
Note that the experimental constraint (\ref{eq:Kbr}) 
applies to an almost massless Goldstone boson.
Since $\pi^+$ with kinetic energy  
larger than 124 MeV were not observed in \cite{Kdecay} ($2 \sigma$), 
the constraint is valid at least for $m_G \lsim 54$ MeV.

When $m_G$ is larger than $2 m_e \simeq 1$ MeV, the rare decay 
$K^+ \rightarrow G + \pi^+$ could be followed 
by $G \rightarrow e^+ e^-$. 
This process through the on-shell Goldstone boson should not 
yield the observed rate\footnote{The on-shell Goldstone process 
has a particular kinematics, so that the constraint on $(F_G,m_G)$ 
should be more stringent than is discussed here. We do not discuss 
this issue further in this article.} \cite{PDG}
\begin{equation}
 {\rm Br}(K^+ \rightarrow \pi^+ e^+ e^-) = 
    (2.88 \pm 0.13) \times 10^{-7}.
\end{equation}
However, for large $F_G$, the decay length (\ref{eq:lengthGee}) is 
so long that, for any range of $m_G$, this condition does not 
yield a limit more stringent than we have already obtained.

\subsection{Beam dump experiment}

Here, we discuss the limits from a beam dump experiment at SLAC  
\cite{SLACbeam}.
An electron beam with energy 12--19 GeV is dumped on a target, 
where the Goldstone boson can be produced 
through a bremsstrahlung-like process.
About 40 $\times (1.6 \times 10^{-19})^{-1}$ electrons were
supplied during the experiment.
A detector is separated from the target by 55 m of dirt, 
and is sensitive to muons. 
Weakly interacting particles such as the Goldstone boson 
can penetrate through the dirt. 
The Goldstone boson can be detected through $\mu^+ \mu^-$ 
pair-creation process. 
Since the muon events were not observed, the couplings of the 
Goldstone boson to electrons and muons have to be sufficiently small.
The number of expected events is obtained by modifying the result 
in \cite{DFLPS} a little:
\begin{equation}
 N = 5.5  \left(\frac{250 \; \GEV}{F_G}\sin^2 \beta\right)^4. 
\end{equation}
This constraint is applied when the Goldstone boson with energy 
of the order of GeV does not decay before it runs 55 m. 
The parameter region excluded by this constraint is shown 
in Fig.~\ref{fig:terr}.

A beam dump experiment at KEK used electron beam and 
$G \rightarrow e^+ e^-$ decay for detection \cite{KEKbeam}. 
The limit on  $F_G$ is improved for $m_G > 2 m_e$. 
The limit for $m_G < 2 m_e$ is improved by 
a beam dump experiment at SIN, which used proton beam 
and $G \rightarrow \gamma\gamma$ decay for detection \cite{SINbeam}. 
The limits from these experiments are shown in Fig.~\ref{fig:terr}.

\subsection{Reactor experiments}

A reactor experiment \cite{ReinesA} is designed to measure the
properties of another weakly interacting particle, the anti-neutrino.
Nuclei in excited states in the reactor decay to states 
with lower energy, emitting $\gamma$ rays. 
But the $\gamma$ ray can be replaced by the Goldstone boson. 
The Goldstone--nucleon coupling (\ref{eq:NNG}) is responsible 
for the emission. 
The flux of the Goldstone can be estimated from $\gamma$ ray spectrum, 
but there is large uncertainty in the estimate.

The flux of the Goldstone boson could have been detected through  
various processes such as $G \rightarrow e^+e^-$ (for $m_G > 2 m_e$), 
$G \rightarrow \gamma \gamma$ and $G+e^- \rightarrow e^- +\gamma$. 
The absence of significant excess in the number of events sets limits 
on the parameter space $(F_G,m_G)$.
Since all the limits obtained from the reactor experiment 
\cite{ReinesA} have been improved by other experiments, however, 
we do not describe this experiment in more details.

Another reactor experiment \cite{ReinesB} is designed to detect 
the anti-neutrino through the neutral current reaction
\begin{equation}
 \bar{\nu} + d \rightarrow p + n + \bar{\nu}.
\end{equation}
The detector is located at a distance of $11.2$ m from the reactor.
The Goldstone boson also induces a similar signal in the detector 
through its axial vector coupling with nucleons.
The expected and observed number of events are \cite{DFLPS} 
\begin{equation}
 4 \times 10^3  \left(\frac{250 \; \GEV}{F_G}\right)^4 
  \; /{\rm day} <  -2.9 \pm 7.2 \; / {\rm day}.
\end{equation}
This constraint is independent of $m_G$ (for $m_G < 2 m_e$).

\begin{figure}[t]
\begin{center}
  \includegraphics[width = .6\linewidth]{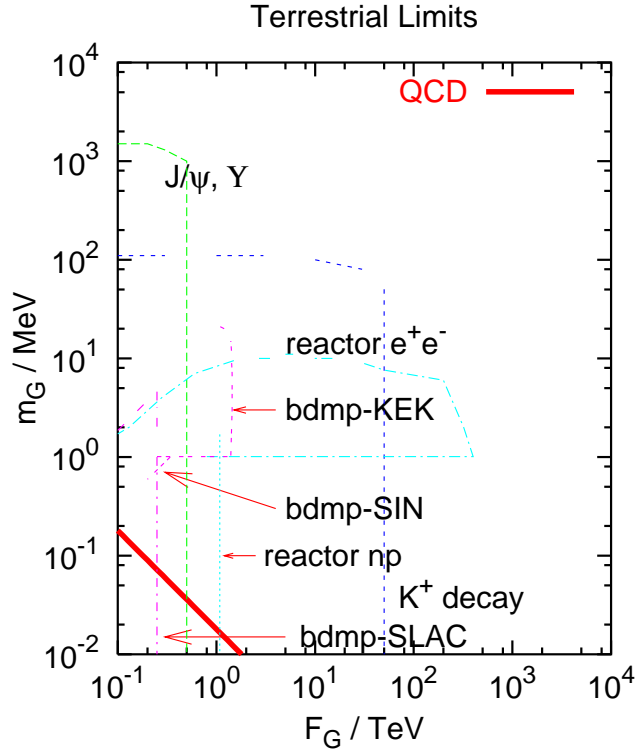} \\
\end{center}
\caption{\label{fig:terr}Various limits from terrestrial experiments 
are schematically described (coloured online). 
``bdmp-SLAC'', ``bdmp-KEK'', and ``bdmp-SIN'' in the figure stand 
for the region excluded by the beam dump experiments at SLAC, KEK, 
and SIN, respectively, and  ``reactor $e^+ e^-$'', and ``reactor $np$''
for that by the reactor experiments through the process 
$G \rightarrow e^+ + e^-$, and $G+d \rightarrow n + p$, respectively. 
The limit ``bdmp-KEK'' is taken from \cite{KEKbeam} and 
``reactor $e^{+}e^{-}$'' from \cite{Bugey}.
Note that the theoretical uncertainties are so large that details 
in this figure do not have importance. 
The limits from LEP experiments $F_G \gsim$ (a few TeV) is not 
shown in this figure.
(The $m_G\mbox{--}F_G$ relation of the QCD axion is shown by 
a thick (red) line.) } 
\end{figure}

An experiment \cite{Bugey} has a detector sensitive to 
$G \rightarrow e^+e^-$ at a distance of $18.5$~m from the reactor 
core, and improves the limit for $m_G>2 m_e$.
$F_G$ has to be large enough so that most of the Goldstone bosons 
pass through the detector without decaying in it, or otherwise, 
$F_G$ has to be small enough so that most of them should have 
decayed before they arrive at the detector. 
The excluded region obtained by \cite{Bugey} is shown 
in Fig.~\ref{fig:terr}.

The Goldstone boson can decay to the lightest neutralinos if
$2 m_{\tilde{\chi}^0_0} < m_G$.
But this possibility does not essentially change the limits.
This is because $m_{\tilde{\chi}^0_0} \lsim$(several MeV) requires
$\lambda \lsim 1/100$, and hence the decay length 
(\ref{eq:lengthGcc}) can never be much shorter than 10 m.

\section{Astrophysical Limits}
\label{sec:astro}

\subsection{The Sun}

The thermal plasma at the core of the Sun produces the Goldstone 
boson through the electron Compton process\footnote{Note that 
the coupling of the Goldstone boson with photons is small  
due to an accidental cancellation (\ref{eq:QED-anomaly}).} 
$e+\gamma \rightarrow e+G$. 
For $F_G \gsim 100 \; \GEV$, the Goldstone boson streams out of the 
Sun without being scattered. 
The energy loss through the Goldstone boson has to be 
at least less than the luminosity of the Sun, 
or otherwise hydrogen in the Sun would have been consumed by now  
\cite{RStr,CS,Raffelt-PRpt1}. 
This condition excludes\footnote{$F_G / \sin^2\beta \lsim 100$ GeV 
is also excluded because the Goldstone boson contributes 
too much to the opacities in the Sun.} 
$100 \; \GEV \lsim F_G / \sin^2 \beta \lsim 1.1 \times 10^{7}$ 
GeV for $m_G \lsim (50 \sim 10)$ keV.

\begin{figure}[ht]
\begin{center}
   \includegraphics[width = .7\linewidth]{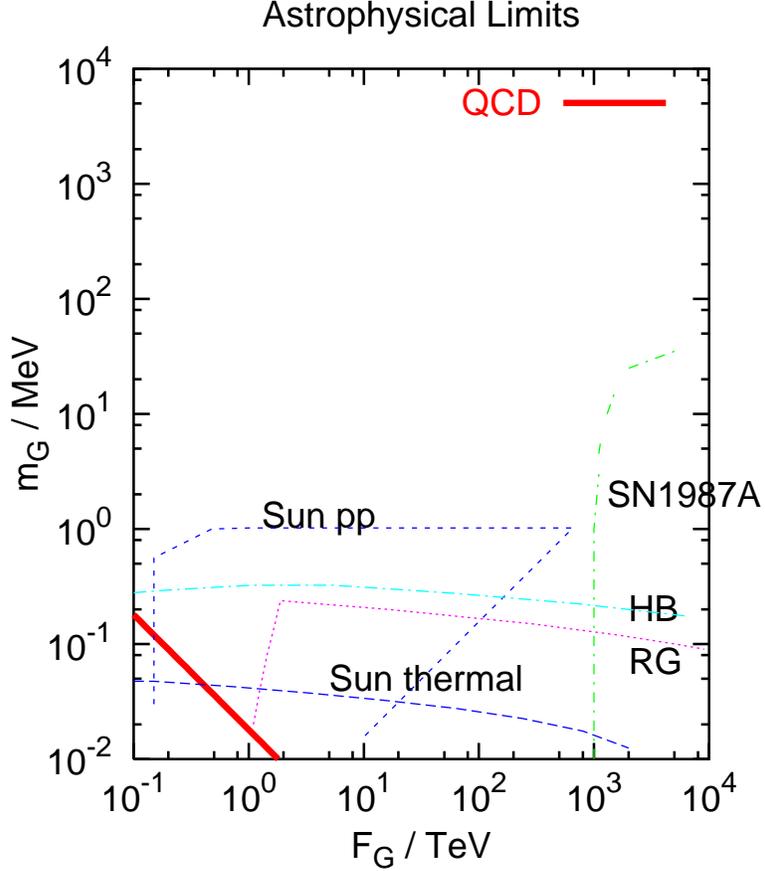} \\
\end{center}
 \caption{\label{fig:astro}Various limits from astrophysics 
(coloured online). The limits from the Sun (emission from the thermal
 plasma) and horizontal branch stars, HB in the figure, are taken from 
\cite{RStr,CS} with $m_G$ increased by 2.5 keV for the Sun and 
by 25 keV for the horizontal branch stars. The limit from red giants, 
RG in the figure, is taken from \cite{Raffelt-PRpt1}. 
The limit from the Goldstone boson emitted from pp-chain nuclei 
does not take account of the possible decay 
$G \rightarrow \tilde{\chi}^0_0\tilde{\chi}^0_0$. 
The limit from SN 1987A for $m_G \gsim 1 \; \MEV$ requires numerical 
analysis. c.f. \cite{MT}.}
\end{figure}

The above requirement is rather conservative. 
A more stringent constraint follows from 
the precise measurement of the solar neutrino flux and 
better understanding of the helioseismology \cite{Raffelt-PRpt2}.
The energy loss should be less than 10 \% of the luminosity of the Sun.
Thus, heavier $m_G$ is excluded; the volume emission rate of the
Goldstone boson becomes $1/10$ times smaller for 
$m_G$ larger by $2\sim 3 \; \KEV$ because of the Boltzmann factor 
$\sim e^{-m_G/(T \sim 1 \; \KEV)}$.
The excluded region, which is essentially the one in \cite{RStr,CS}, 
is shown in Fig.~\ref{fig:astro} (Sun thermal).

Although the emission from the thermal plasma $(T \sim 1 \; \KEV)$ 
is suppressed for $m_G \gsim 50 \; \KEV$, it is possible to emit 
Goldstone bosons in nuclear processes \cite{RStd}. 
For instance, the p-p chain involves a process 
$p+d \rightarrow {}^3{\rm He}+\gamma(5.5 \; \MEV)$, and 
the Goldstone boson can be emitted instead of the 5.5 MeV $\gamma$ ray 
as long as $m_G \ll 5.5 \; \MEV$.
The emission rate is given by 
\begin{equation}
  \left[\frac{1}{2}
  \frac{(F^{(1)}_A)^2 \frac{\left(m_p/F_G \right)^2}{4\pi}}
       {\alpha_{QED}}
  \left(\frac{\rho^{(1)}}{4.7}\right)^2 \right]
  \times 1.7 \times 10^{38} /{\rm sec,}
\label{eq:Gflux-Sun55}
\end{equation}
where $m_p$ denotes the mass of proton, and $F_A^{(1)}$ 
and $\rho^{(1)}$ are defined in Eq.~(\ref{eq:isovect-form}). 
Since the factor in the bracket is much smaller than unity, 
this flux does not have significant effects on the evolution 
(through energy loss) or structure (through opacity) of the Sun. 

Once created at the centre of the Sun, if the Goldstone bosons are 
absorbed by scattering or disappear through decay within the 
radius of the Sun,   
then there is no phenomenological constraint.
This is the case for $F_G \lsim 100 \; \GEV$, 
when the electron Compton process $G + e^{-} \rightarrow e^- + \gamma$ 
absorbs the 5.5 MeV Goldstone boson 
while it is inside the Sun \cite{RStr}; 
only the axial vector coupling $1/F_G$ is relevant here, and not $m_G$.
As $m_G$ becomes larger, the decay process 
$G \rightarrow 2\gamma$ becomes more important than the electron 
Compton scattering. 
The decay length becomes shorter as $m_G^{-4}$, 
and the 5.5 MeV Goldstone boson can decay 
within the radius of the Sun ($R_{\odot} \sim 6 \times 10^8$ m)
for sufficiently large $m_G$, even if $F_G \gsim 100 \; \GEV$.
As $m_G$ becomes larger than $2 m_e$, 
the Goldstone boson decays to $e^+e^-$ within the Sun, 
as long as $F_G \lsim 10^3 \; \TEV$.
For some parameter region, the Goldstone boson decays also 
to $\tilde{\chi}^0_0 \tilde{\chi}^0_0$ within the Sun, and 
the flux does not come out of the Sun as the Goldstone bosons.

If none of the above processes succeeds in trapping the Goldstones 
inside the Sun, they will escape from the Sun towards the Earth.
If the Goldstone boson cannot decay either to $e^+ e^-$ or to 
two $\tilde{\chi}^0_0$'s, then they will decay to two photons.
Among the Goldstone bosons from the Sun, 
only a fraction $1 {\rm AU}/(l_{2\gamma} \gamma)$ are 
observed as $\gamma$ rays on the Earth, where 
$1 {\rm AU} \simeq 1.5 \times 10^{11}{\rm m}$ is 
the distance between the Sun and the Earth.
The $\gamma$-ray flux from the Sun is 
less than $0.8 \times 10^{-3} \; \MEV /{\rm cm^2~sec} $ \cite{RStd} 
for $\gamma$-ray energies between 4 MeV and 6 MeV.
This means\footnote{If $m_G/(5.5 \; \MEV)$ 
is not negligible, the $\gamma$-ray spectrum is not the same 
as the one assumed in \cite{RStd}. We do not discuss this issue.} 
that fewer than $6.2 \times 10^{24}/${\rm sec} Goldstone bosons 
from the Sun are observed as $\gamma$ rays \cite{RStd}.
Since the flux of Goldstone bosons (\ref{eq:Gflux-Sun55}) is 
much larger than this observational bound, only a small fraction of 
the Goldstone flux can decay before reaching the Earth.
The decay length (\ref{eq:lengthGgg}) has to be much larger 
than $1 {\rm AU}$. 
The region excluded by this $\gamma$-ray observation 
is shown in Fig.~\ref{fig:astro} (Sun pp), where the decay 
$G \rightarrow \tilde{\chi}^0_0 \tilde{\chi}^0_0$ is not taken 
into account.

For sufficiently large $F_G$, the flux of the Goldstone boson 
can arrive at the Earth without decaying to photons.
The solar axion search tries to detect the flux 
by converting the Goldstone boson into photon in the presence 
of strong magnetic field; 
the conversion is due to the Goldstone boson--photon mixing 
that arises from the coupling (\ref{eq:G-QED}). 
The conversion rate is given by the square of the mixing angle.  
When we consider the flux of 5.5 MeV Goldstone bosons, 
the conversion probability is of order 
\begin{equation}
 \left(0.113 \times \frac{\alpha_{QED}}{\pi F_G}B \right)^2 
 \left(\frac{5.5 \; \MEV}{m_G^2}\right)^2 \sim 
 6 \times 10^{-27} \left(\frac{B}{10 {\rm T}}\right)^2 
 \left(\frac{10^2 \; \TEV}{F_G}\right)^2 
 \left(\frac{100 \; \KEV}{m_G}\right)^4.
\end{equation}
Since the emission rate (\ref{eq:Gflux-Sun55}) corresponds to 
a flux $\left[ \cdots \right] \times 6.0 \times 10^{10}{\rm /cm^2 s}$ 
at the Earth, 
the conversion rate is too small for a signal to be detected.
An axion search experiments like this is mainly sensitive to light 
pseudo-scalar particles, because the conversion rate is highly 
suppressed for large $m_G$.

\subsection{Red Giants and Helium-Burning (Horizontal-Branch) Stars}

As hydrogen is burnt inside stars, helium is accumulated at the core 
and compressed.
The core of low-mass stars becomes high-density and degenerate. 
The thermal plasma in the core produces the Goldstone boson 
dominantly through the bremsstrahlung process 
$e^- + ({\rm nucleus}) \rightarrow ({\rm nucleus}) + e^- + G$. 
This process dominates over the electron Compton process partly 
because the number density of nucleons becomes much larger than 
that of photons in the high-density core.
When the Goldstone boson carries away too much energy from the core, 
the core is cooled and the helium-burning process 
$3 {}^4 \! {\rm He} \rightarrow {}^{12}\! {\rm C}$ 
is not ignited until a later time \cite{Raffelt-PRpt2}.

The Goldstone boson streams out of the helium core, whose radius 
is about $10^7$m, 
if $ F_G /\sin^2 \beta \gsim 0.9 \times 10^3 \; \GEV $.
In that case, the number of produced Goldstone boson has 
to be sufficiently small. 
$F_G /\sin^2 \beta \lsim 2 \times 10^9 \; \GEV$ is excluded 
for $m_G \lsim (100 \sim 200)\; \KEV$ \cite{Raffelt-PRpt1}.

After helium starts burning, excessive energy loss through 
free-streaming Goldstone bosons leads to excessive consumption 
of helium, shortening the lifetime of such stars 
(called horizontal-branch stars).
The Goldstone boson streams out of such stars, 
without being scattered by the electron Compton process, 
if $F_G \gsim 10^3\mbox{--}10^4 \; \GEV$. 
In this case, the energy loss has to be at least less than 
the luminosity. Thus, $F_G / \sin^2 \beta \lsim 5 \times 10^8$ GeV is 
excluded for $m_G \lsim 300 \; \KEV$ \cite{RStr,CS}. 
Recent articles conclude \cite{Book,Raffelt-PRpt2} 
that the energy loss should be less than 10\% 
of the nuclear energy release, and hence 
the lower bound on $m_G$ is pushed up by $20\mbox{--}30 \; \KEV$, 
as in the case of the energy-loss argument for the Sun.
A similar lower bound on $m_G$ follows for 
$F_G\lsim 10^{3}\mbox{--}10^4 \; \GEV$ because the Goldstone boson 
should not contribute too much to the heat transfer rate 
inside the horizontal-branch stars \cite{RStr,CS}. 

The parameter region excluded 
by the constraints from red giants and horizontal-branch stars
is shown in Fig.~\ref{fig:astro}.
Even when the Goldstone boson can decay to two $\tilde{\chi}^0_0$, 
the energy is lost anyway, 
so that, the excluded region does not change very much 
(except for small $F_G$, which is not our main concern).

\subsection{White Dwarfs}

After helium is burnt, light stars become white dwarfs. 
The cooling rate of white dwarfs near the solar system has been 
measured, and hence the rate of energy loss through 
Goldstone-boson emission is constrained from above.
The lower bound on $F_G$ obtained in this way is similar 
to the one obtained from red giants and horizontal-branch stars 
\cite{Raffelt-PRpt1}.
Since the temperature of the white dwarfs is less than that of 
horizontal-branch stars, the lower bound on $m_G$ is not strengthened.

\subsection{Supernova 1987A}

Large-mass stars experience supernova explosions after 
carbon, oxygen and other heavy elements are burnt.
The case of supernova 1987A allowed 
various observations of the explosion which set limits on 
possible new physics. 

The Goldstone boson $G$ would have been produced through the nucleon 
bremsstrahlung process $N+N \rightarrow N+N+G$ 
in the collapsing iron core, 
as long as $m_G$ is less than (a few)$\times 10 \; \MEV$. 
But the density at the core is so high that the inverse process 
$N+N+G \rightarrow N+N$ can absorb the Goldstone bosons. 
They stream out of the supernova from a constant-radius sphere 
where the density becomes sufficiently small that the optical depth 
becomes of order unity. 
As the nucleon--Goldstone boson coupling $m_N/F_G$ becomes large,
the Goldstone-boson emitting surface goes outward, the temperature 
at the surface decreases, and 
the energy loss through the Goldstone-boson flux decreases.
Only $F_G \lsim 10^3$ TeV is allowed \cite{Raffelt-PRpt1}. 

Although the flux of the Goldstone boson decreases as $F_G$ becomes
smaller, the Goldstone boson--nucleon interaction cross section 
increases. 
If the Goldstone boson were to arrive at the Earth from the supernova 
1987A, the flux would roughly be 
\begin{equation}
 \phi_{G,\oplus} \sim 
    \left(\frac{F_G}{10^3 \; \TEV}\right)^{\frac{36}{49}} \times 
    \left[\phi_{\bar{\nu}_e} \sim 10^{10} {\rm /cm}^2  \right],
\label{eq:Gflux-SN}
\end{equation} 
where a model of supernovae $T \propto$ (number density)$^{1/3}$ 
and (number density)$\propto r^{-p}$ with $p=5$ is assumed  
\cite{nuclearKII}.
Since the cross section of nuclear excitation is proportional to 
$1/F_G^2$,  the expected number of events in the detectors of 
Kamiokande II and IMB would have been increased for small $F_G$.
Thus, a certain parameter region would have been excluded because 
of the absence of such events \cite{nuclearKII}. 
However, for the parameter region ($F_G \lsim 10^3$ TeV, 
$m_G \gsim 300 \; \KEV$), which is not excluded either 
by the energy-drain from the supernova 1987A or 
by the helium consumption of the horizontal-branch stars,  
the Goldstone boson decays into two $\gamma$ rays 
within $10^{18}$ m $\sim 30$ pc.  
Therefore, the Goldstone boson did not arrive at the Earth, and 
the above constraint does not exclude any parameter space.
Instead, the decay product might have been observed.

If $m_G$ is larger than $2 m_e$, the Goldstone boson decays to 
$e^+ e^-$. The decay length is at most $10^8$ m 
for $F_G \lsim 10^3$ TeV and $m_G \gsim 1 \; \MEV$, and hence is 
much smaller than the radius of the star ($\sim 10^{11}$ m) 
before the explosion.\footnote{This is not the radius of the core, 
but the radius of the star including the hydrogen and helium shell.}
Thus, the decay products are absorbed inside the star, and are not 
observed from outside.\footnote{\label{fn:SNenergetics} 
Even if the decay length is smaller than the radius, 
the Goldstone-boson flux releases the energy of the gravitational 
collapse at the mantle and/or envelope through energetic $e^+ e^-$. 
On the other hand, the observation of supernova 1987A confirmed 
that the gravitational energy is released mainly through neutrino 
emission, and not through the explosion of the mantle and envelope. 
Thus, excessive energy transfer through the Goldstone-boson flux 
contradicts observation.
In particular, the decay length has to be sufficiently short. 
It should be less than $10^{11}$m \cite{Raffelt-PRpt2}, 
but the precise upper bound is not clear. 
We do not discuss this issue further.}

If $m_G$ is less than $2 m_e$, the Goldstone boson decays 
to two $\gamma$ rays. 
The decay length is larger than $10^{14}$ m for 
the allowed parameter space with $m_G < 1$ MeV, 
and hence the $\gamma$ rays will not be absorbed by the supernova 
itself, whose radius was about $10^{11}$m. 
The $\gamma$ ray should have been observed for this parameter region, 
but significant excess of the counts of $\gamma$ rays was 
not observed in the range of $4.1\mbox{--}6.4$ MeV \cite{Satellites}.
The observational upper bound on the $\gamma$-ray fluence 
in this energy range is $\phi_{\gamma,\oplus} \lsim 0.9/${\rm cm}$^2$, 
which is much smaller than (\ref{eq:Gflux-SN}).
Thus, the parameter region with $m_G < 2m_e$ is allowed only if 
$m_G > 2 m_{\tilde{\chi}^0_0}$, and the Goldstone boson decays 
dominantly to the LSP, rather than to $\gamma$ rays.

When kinematics allows the Goldstone boson to decay to the LSP, 
it generically decays dominantly to the LSP rather than $\gamma$ rays;
this is not an additional assumption, but can be seen by comparing 
(\ref{eq:lengthGgg}) and (\ref{eq:lengthGcc}).
Thus, the flux of LSP arrives at the Earth, instead of the Goldstone 
bosons.
But the LSP's are not detectable because of their small interaction 
cross section.
The LSP flux does not contribute to the energy transfer 
to the mantle and/or envelope of the supernova, either, because 
of the small scattering cross section.  
Therefore, no limits come from the flux of the LSP.
However, some of the Goldstone bosons still decay 
to two $\gamma$ rays, and the $\gamma$ rays could have been observed.
We postpone the discussion on this issue to section \ref{ssec:DM-SN}.

\noindent
{\bf Summary of Phenomenological Limits}\\
Here, we briefly summarize the parameter region 
allowed by the various phenomenological limits.
For $m_G$ larger than 100 MeV, $F_G \gsim$ (a few TeV) is allowed; 
the lower bound on $F_G$ comes from the LEP experiment.
For $300 \; \KEV \lsim m_G \lsim ({\rm several}) \times 10 \; \MEV $, 
the allowed region is $10^{2} \; \TEV \lsim F_G \lsim 10^{3} \; \TEV$.
The upper bound comes from excessive energy loss from SN 1987A, and 
the lower bound from the reactor experiment 
(for $2m_e < m_G <$(several MeV)), the Goldstone emission 
from the pp chain (for $m_G < 2m_e$), 
and from $K^+$ decay. 
The parameter space with $m_G < 2m_e$ is allowed only when 
$m_G > 2m_{\tilde{\chi}^0_0}$.
The parameter space with $m_G \lsim 300 \; \KEV$ is excluded by 
energy loss from horizontal-branch stars.

\section{Cosmology}

\subsection{The BBN Era}

In our theory, cosmology below the electroweak scale differs
significantly from that of the MSSM because the electroweak sector
contains three light states. There is the Goldstone, $G$, with mass
$m_G$ and decay constant $F_G$; the scalar $H_1$, with mass $\approx
\lambda v / \tan \beta$; and the LSP, $\tilde{s}$, with
mass $\approx \lambda^2 v$. 
The cosmological behaviour of these states is
dependent on the parameters ($F_G, m_G$), and although this parameter
space is highly constrained from both terrestrial and astrophysical
arguments, as shown in Figures 1 and 2, wide regions still remain to
be explored. 
In this section we will restrict our attention to 
$m_G \lsim ({\rm several})\times 10$ MeV,
in which case $F_G \lsim 10^3$ TeV due to the 
limit from SN 1987A.
In this case, $H_1$ is heavier than about 10 MeV, and 
sufficiently strongly coupled that, as the temperature drops 
below its mass, it decays to $e^+ e^-$ or $\tilde{s} \tilde{s}$. 
It leaves no cosmological signal, and we consider it no further.

The reaction $\gamma e \rightarrow G e$ has a rate of order 
$ \alpha m_e^2 T/ F_G^2$ and recouples $G$ to the $e/\gamma$ fluid 
before BBN at a recoupling temperature
\beq
T_R \approx \mbox{GeV}  \left( \frac{50 \mbox{TeV}}{F_G} \right)^2,
\label{eq:TR}
\eeq
provided $T_R > m_G$. 
An important question is whether $G$ is lighter or heavier than the
MeV scale. If $m_G < $ O(MeV), then $G$ inevitably has an effect 
on BBN. 
On the other hand, if $m_G \gg$ O(MeV), then $G$ will decay 
to $e^+ e^-$, with a rate given by (\ref{eq:Gtoee}). 
For most of this parameter range, these decays are in equilibrium 
as the temperature drops below $m_G$, so there is no effect on BBN. 
Even in the case when the decays occur at a lower temperature 
than $m_G$, the reheat temperature is always above the MeV scale. 
Hence, for any $m_G \gg$ O(MeV) and $F_G \lsim 10^3$ TeV, 
$G$ does not affect BBN.

We now concentrate on the range $300 \; \mbox{keV} \lsim m_G \lsim 3
\; \mbox{MeV}$, that leads to non-standard effects in BBN. We do not
attempt a numerical analysis of the BBN era, but discuss qualitative
features. There are several important temperature scales that are
close to each other, so that even the qualitative picture depends
sensitively on $m_G$. For simplicity we assume that $\nu_{e,\mu,\tau}$
all decouple from the $e/\gamma$ fluid at the same temperature, $T_\nu
\approx 3$ MeV. (The $\nu_{\mu,\tau}$ decouple first at 3.7 MeV, 
followed by the $\nu_e$ at 2.4 MeV.) 
The neutron to proton ratio freezes out at $T_{np} \approx 0.8$ MeV, 
and we assume that $N_{\nu BBN}$ is dominantly 
determined by the radiation energy density at this era.\footnote{
The neutron to proton ratio weakly depends on the time of 
deuterium formation. Although the Goldstone boson and its annihilation 
process affects BBN through this time scale, 
we neglect this effect, and discuss only the dominant effect.}  
At $T \approx m_e$, electron-positron annihilation heats the photon
fluid to a temperature above that of the neutrinos. 

As we have seen, the reaction  $\gamma e \rightarrow G e$ recouples
before BBN, hence if $m_G \lsim T_{np}$ the $G$ will be present with a
full thermal abundance during $n/p$ freezeout, so that we expect
$N_{\nu BBN} \simeq 3 + 4/7$. 
This is still consistent with the observed abundance of 
light elements ($2\sigma$).
The situation is more complicated if $T_{np} \lsim m_G \lsim T_\nu$. 
The reactions  $G e \rightarrow \gamma e$ and $G \rightarrow e^+ e^-$ 
lead to an exponential decrease in the number density of $G$ 
as the temperature drops below $m_G$, heating the $e/\gamma$ 
fluid relative to the decoupled neutrinos. 
(We argue later that $G \rightarrow \tilde{s} \tilde{s}$ must 
not be in thermal equilibrium at this era). 
This effect leads to $N_{\nu BBN} \simeq 3(11/13)^{4/3} \simeq 2.4$, 
which is close to the central value inferred from observation. 
However, a significant number of Goldstone bosons are still 
in the plasma when $m_G$ is not much larger than $T_{np}$, and the 
energy density of such Goldstone boson contributes to $N_{\nu BBN}$.
Furthermore, the interaction rates for $n \nu \leftrightarrow  p  e^-$ 
and $n e^+ \leftrightarrow p  \bar{\nu}$ differ from the standard case,
because of the decreased number density and average energy of the
neutrinos, so that a careful analysis is needed to determine the shift
of $N_{\nu BBN}$ from 2.4. 

\subsection{Signals in the Cosmic Microwave Background}

The acoustic oscillations during the eV era leave an imprint on the
cosmic microwave background, allowing a determination
of the total radiation energy density during that era, often
parameterized as $N_{\nu CMB}$. Since $H_1$ decays well before BBN and
$\tilde{s}$ are non-relativistic by this era (assuming they are stable),
the only possible $N_{\nu CMB}$ signal would arise from $G$. If
$m_G \gsim T_\nu$, the temperature of neutrino decoupling, then the
removal of $G$ from the bath heats the $e/\gamma / \nu$ fluid equally:
as with BBN, there is no signal in this mass range. However, if $m_G
\lsim T_\nu$, then only the $e/ \gamma$ are heated, leading to the
prediction
\beq
N_{\nu CMB} = 3 \left( \frac{11}{13} \right)^{4/3} = 2.40.
\label{eq:nvcmb}
\eeq
This is a remarkable signal: many other phenomena lead to 
$N_{\nu CMB} > 3$ \cite{cmbsig}, 
but $N_{\nu CMB} < 3$ can also be realized in a simple way: 
a light scalar particle is in equilibrium when $T \approx$ 3 MeV, 
and decays before deuterium formation, 
heating the $\gamma$ plasma relative to neutrinos.
For alternative origins for $N_{\nu CMB} < 3$, see e.g. 
references cited in \cite{BBNCMB}.

\subsection{LSP $\tilde{s}$ Dark Matter}

If $R$ parity is conserved, the lightest superpartner 
$\tilde{\chi}^0_0$, or almost equivalently $\tilde{s}$, is
stable and could be the cosmological dark matter. 
It has a mass much less than the usual LSP candidates. 
Limits on warm dark matter from WMAP reionization data 
is $m_{\tilde{s}} \gsim 10$ keV \cite{warm-reion}. 
This constraint is roughly satisfied by $F_G \lsim 10^3$ GeV.

To give $\Omega_{\tilde{s}} \simeq 0.2$, we require 
\begin{equation}
  \frac{n_{\tilde{s}}}{n_\gamma}
  \left(\frac{m_{\tilde{s}}}{100 \; \KEV}\right) 
  \approx 0.3 \times 10^{-4}.
\end{equation}
A thermal abundance of stable $\tilde{s}$ would exclude our
theory. Since $\tilde{s}$ were in thermal equilibrium at the
electroweak scale, we require that they decoupled from the $e/ \gamma$
fluid before BBN, and entropy generation (for example from the QCD
phase transition or from late decaying non-relativistic particles)
depleted $n_{\tilde{s}}$ by a factor of order 
$10^{4\sim 5} (m_{\tilde{s}}/100 \; \KEV)$. 
Direct interactions of $\tilde{s}$ with the $e/ \gamma$
fluid do indeed decouple well before BBN. The reaction $e^+ e^-
\leftrightarrow \tilde{s}\tilde{s}$ decouples at a temperature of 10
GeV for $F_G \simeq 100$ TeV, leaving four orders of magnitude
in temperature for entropy generation to occur before BBN. Since $G$
is thermally coupled to $e/ \gamma$, we must also ensure that $G
\leftrightarrow \tilde{s}\tilde{s}$ also decouples before the BBN
era. For large $m_G$ and $F_G \sim 10^3$ TeV, this is immediate, 
but for values of $m_G$ of order the MeV scale or below, 
further analysis is necessary. 

\subsection{$m_G \lsim 2 m_e$: Consistency between SN 1987A and Cosmology?}
\label{ssec:DM-SN}

If the reaction $G \leftrightarrow \tilde{s} \tilde{s}$ is in thermal 
equilibrium during the cosmological era with temperature 
$T \simeq m_G$, a full thermal abundance of
$ \tilde{s}$ will be created. Dilution of these $ \tilde{s}$ by
entropy production is not possible after BBN, because the BBN value of
$n_B/n_\gamma$ is consistent with the values from CMB and today. 
One way to avoid overclosing the universe from these
cosmologically produced $\tilde{s}$ is for them to decay before they dominate the
universe. Obtaining such a large decay rate, whether to a light gravitino
or via $R$ parity breaking, appears difficult. To avoid overclosure by
a stable $\tilde{s}$, we must limit the $\tilde{s}$ production:
$n_{\tilde{s}} / n_\gamma \approx \Gamma(G \rightarrow \tilde{s}
\tilde{s}) \; t( T = m_G) \; < \; 10^{-4\sim -5} 
(100 \; \mbox{keV} / m_{\tilde{s}})$, implying
\beq
\Gamma(G \rightarrow \tilde{s} \tilde{s}) < 
\frac{1}{5 \times 10^{12} \; \mbox{m}} 
\left( \frac{100 \;  \mbox{keV}}{m_{\tilde{s}}}  \right) 
\left( \frac{m_G}{\mbox{MeV}} \right)^2.
\label{eq:sinonumber}
\eeq

A large flux of $G$ was emitted from SN 1987A. 
If the only open decay channel is $G \rightarrow \gamma \gamma$, 
our theory would be excluded by the non-observation of $\gamma$ rays 
coincident in time with the observed neutrino burst. 
Hence if the $e^+ e^-$ channel is closed, we must require 
the $G \rightarrow \tilde{s} \tilde{s}$ channel be open.
Using the cosmological limit (\ref{eq:sinonumber}) gives a lower bound
to the $\gamma \gamma$ branching ratio:
\beq
B_{\gamma \gamma} = \frac{ \Gamma(G \rightarrow \gamma \gamma)}{\Gamma(
G \rightarrow \tilde{s} \tilde{s})} > 10^{-2}
 \left( \frac{m_{\tilde{s}}}{100 \; \mbox{keV}}  
 \right) \left( \frac{m_G}{\mbox{MeV}} \right)
 \left( \frac{10^3 \; \mbox{TeV}}{F_G} \right)^2.
\label{eq:Bgg}
\eeq
If $G$ escape from the progenitor star, such a large $B_{\gamma
  \gamma}$ implies a flux of $\gamma$ rays from SN 1987A that would
have been detected on earth. We must require that
$ \Gamma(G \rightarrow \tilde{s} \tilde{s}) > 1/L_\gamma$, where
$L_\gamma$ is the distance from the supernova core to the radius at
which MeV $\gamma$ rays can escape from the progenitor star. Comparing
with the cosmological limit (\ref{eq:sinonumber}), we require
that $L_\gamma$ be larger than the inverse of the right-hand side 
of (\ref{eq:sinonumber}). 
Since this is close to the radius of the progenitor star of SN 1987A, 
$10^{11}$m, a more detailed calculation is required 
to determine whether $m_G < 2 m_e$ is allowed. 
If this region is found to be allowed, then we note
that the cosmological bound (\ref{eq:sinonumber}) is close to being
saturated. Hence it may be that the $n_{\tilde{s}}$ is
first diluted by a very large amount, and then brought back to 
an appropriate order by the $G \rightarrow \tilde{s} \tilde{s}$ 
process. 

\subsection{Topological Defects} 

After the electroweak phase transition, the Peccei--Quinn symmetry 
is spontaneously broken, and global cosmic strings are formed. 
But the energy density of the strings is very small, and there is 
no significant impact on the density perturbation.

As the temperature drops further, the mass of the Goldstone boson 
becomes important, and domain walls bounded by strings are formed. 
If there is only one vacuum, the walls and strings shrink 
and eventually vanish. 
Otherwise, the energy density of the walls dominates the universe, 
and causes a cosmological problem. 
The number of vacua depends on how the mass of the Goldstone boson 
is generated. 
Since we assume that the mass is due to explicit breaking of the 
symmetry at high-energy scale, domain walls are not a problem 
of the low-energy effective theory, but are an issue for model building 
at high energy.

In case the mass of the Goldstone boson is due to the dimension-5 
operator $[S^4/M]_F$, the $Z_4$ PQ symmetry is spontaneously broken, 
and there are four distinct vacua. 
However, the mod 4 PQ symmetry is not an exact symmetry, 
but is broken by the QCD anomaly, and hence the four vacua are not 
degenerate.
There are no degenerate vacua either when the U(1) PQ symmetry 
is broken by $[\zeta S + \kappa S^3]_F$.

\section{Signals at Future Colliders}
\label{sec:}

From an experimental point of view, our theory differs from the MSSM 
in two important ways. Firstly the $\mu$ parameter is not a free
parameter, but is determined by the pseudoscalar mass $M_A$ and 
$\tan \beta$ (\ref{eq:mupred}). 
Secondly, there are light states $G, H_1, \tilde{\chi}_0$ that
lie dominantly in the singlet superfield, $S$, and have interactions
proportional to the small coupling $\lambda$. These small couplings
imply that the rates for direct production of these states at
colliders will be very small. Hence only the superpartners and Higgs 
states of the MSSM will be directly produced, and furthermore, 
for any given point in parameter space, the production rates will be 
identical to those of the MSSM. Of course the point in parameter space
is now constrained by the prediction for $\mu$. 
The question then becomes: do the cascade decays of superpartners
and Higgs bosons lead to different signals in our theory compared to
the MSSM? 

If $\epsilon$ is as small as $10^{-3}$, then the scalar states $G$ and
$H_1$
are unlikely to be produced in the cascade decays sufficiently often
to yield an observable signal. For example, the decay rate of  
$H_2 \rightarrow GG$ is suppressed by $\epsilon^4$, 
and is generically smaller than that of $H_2 \rightarrow 
\gamma \gamma$. However, for $\epsilon \simeq 0.1$ there will be
spectacular events with  $H_2 \rightarrow GG \rightarrow
e^+e^-e^+e^-$. 

The fermion $\tilde{\chi}_0$ is much more important 
since it is the LSP.\footnote{In the case of gauge mediation, the LSP 
could be the gravitino $\psi_{3/2}$. However, 
the decay process $\tilde{\chi}_0 \rightarrow G \psi_{3/2}$ is not
always allowed kinematically. Even when the decay is kinematically
possible, the decay product $G$ will not be observed, because, 
as long as $m_G \lsim 100$ MeV, the decay length of $G$ 
is much larger than the typical size of the detectors. 
Thus, the $\tilde{\chi}_0$ LSP and gravitino LSP do not make 
a difference in collider experiments.}
Pair production of superpartners will always lead
to final states containing two $\tilde{\chi}_0$.
Note that all the superparticles except the lightest superparticle 
of the MSSM, i.e., the next-to-LSP (NLSP), do not decay 
to the LSP $\tilde{\chi_0}$ because the branching ratio 
is at most of order $\epsilon^2$. 
Thus, $\tilde{\chi}_0$ are emitted only through the decay of the NLSP.
An immediate consequence is that one extra decay process is 
always involved in the cascade decay leading to the LSP, 
and thus the missing (transverse) energy is generically 
degraded relative to the observed (transverse) energy.

Suppose that the NLSP is kinematically allowed to decay 
to a standard model particle $X$ and $\tilde{\chi}_0$, 
${\rm NLSP} \rightarrow X \tilde{\chi}_0$. 
The particle $X$ is usually the ``superpartner'' of the NLSP. 
In our theory, two $X$'s are always emitted in supersymmetric events.
When the NLSP is a neutralino, $X$, the ``superpartner'' of the NLSP, 
is either a $Z$ boson or a scalar Higgs boson $h$.
Since the NLSP does not have to be neutral, the ``superpartner'' $X$  
can be $\tau$ when the NLSP is $\tilde{\tau}$, or a $W$ boson when 
the NLSP is a chargino.
The $\tilde{\tau}_R$ NLSP will be interesting, e.g. in the context 
of gaugino mediation. 

For example, in the
MSSM with a neutralino LSP, squark production at a hadron collider
leads to dijet events with large missing transverse energy, which 
are sometimes accompanied by leptons or more jets. 
In our theory, the same parameter region would lead to events 
with two extra $Z/h$ bosons and a reduced missing transverse energy. 
If the decays of the neutralino to $Z$ dominate over the decays 
to Higgs, approximately 1/3\% of all superpartner pair production 
events would have both $Z$ bosons decaying to either $e^+e^-$ or 
to $\mu^+ \mu^-$. 
When the NLSP is a neutralino $\tilde{\chi}^0_1$ that is lighter than 
the $Z$ boson, the NLSP will dominantly undergo three-body decay 
via a virtual $Z$-boson, but a certain fraction of the NLSP 
may go through a three-body decay to $l^+l^-\tilde{\chi}_0$ 
via a virtual $\tilde{l}$.

At an $e^+ e^-$ collider, pair production of the stable neutral LSP 
of the MSSM e.g., $\tilde{\chi}^0_1$ does not give an observable 
signal. 
However, in our theory, the NLSP pair production is observable through
the decay process to the LSP, and will be extremely interesting. 
In particular, the production cross section and the branching ratio 
of the NLSP directly reveal various properties of the NLSP, 
the LSP of the MSSM.

\section{Conclusions}
\label{sec:conc}
Any supersymmetric extension of the standard model must address 
the questions of how the chargino and axion masses are generated. 
In the MSSM this is accomplished via the superpotential term 
$\mu H_1 H_2$ and the soft term $\mu B h_1 h_2$, 
leading to the $\mu$ problem: why is the supersymmetric mass parameter 
$\mu$ of order the supersymmetry breaking scale? 
We have introduced an alternative highly predictive framework 
for studying supersymmetry at the weak scale, which incorporates the 
Ciafaloni--Pomarol mechanism for the dynamical determination 
of $\mu$ \cite{CP}. 
We assume that the chargino mass is generated via a singlet vev 
in the superpotential term $\lambda S H_1 H_2$, 
but take an agnostic view as to the origin of the axion mass. 
We assume that the explicit PQ symmetry breaking in the effective theory 
at the TeV scale is sufficient to make the axion heavy, 
but does not significantly affect the physics of the electroweak
symmetry breaking at the TeV scale. 
The advantage of this viewpoint is clear: it separates the issues 
of electroweak symmetry breaking and PQ symmetry breaking. 
If this viewpoint is correct, we do not need to understand the origin of 
the axion mass in order to have a theory of electroweak symmetry breaking.
We have given three explicit examples of models that lead to our framework.

Is the NMSSM, with PQ symmetry broken by the superpotential term $\kappa
S^3$,  an example of a model that falls into our class of theories? If
$\kappa$ is of order unity the answer is no --- there are additional
parameters that affect the TeV scale physics of electroweak symmetry
breaking beyond those of our framework. However, in the limit
that $\kappa/\lambda \rightarrow 0$, the NMSSM does become an example 
of our framework, as the interactions leading to $m_G$ are too small
to affect potential minimization and collider physics.  
Alternatively, $m_G$ may be generated from higher dimensional operators 
from physics far beyond the TeV scale. 

Our framework of supersymmetry with an approximate PQ symmetry solves the
$\mu$ problem in a different way than the NMSSM. In the NMSSM 
all dimensionless parameters are of order unity so that the singlet vev, 
and therefore the induced $\mu$ parameter, must be of order 
the supersymmetry breaking scale. In this framework, even if
$\lambda \ll 1$, the minimization equations set $\mu \approx A/ 
\tan \beta$ (provided $m_S^2$ is small enough) for {\em any} $\lambda$ 
\cite{CP}. 

We are interested in the case that the singlet vev is in the (multi-)
TeV domain, and not at the scale of $10^{10}\mbox{--}10^{12}$ GeV
required for invisible axion models, where the axion mass comes solely
from the QCD anomaly. Indeed, the light pseudo-scalar $G$ is too heavy 
to solve the strong CP problem and we should not call it the axion. 
However, we have concentrated on the possibility that $G$ is much
lighter than the weak scale, since it is in this limit that our
effective theory approach becomes accurately valid. We have found
acceptable regions of parameter space with $300 \; \KEV < m_G <
100$ MeV and $100 \; \TEV < \sqrt{2} \vev{s} < 1000$ TeV. 
This requires $\lambda \approx 10^{-3}$, and $|m_S^2 | < \lambda^2 v^2$. 
For heavier $m_G$, the singlet vev could be as low as a few TeV, 
and $\lambda \approx 0.1$. 
Lower values of $m_G$ and/or $\vev{s}$ are excluded by a variety of
terrestrial and astrophysical processes, as shown in Figures 1 and 2. 
In the allowed regions, our theory is essentially the MSSM together 
with a light decoupled singlet superfield.

In the limit of small $\lambda$, the two Higgs doublets are remarkably 
similar to those of the MSSM. This is not surprising since this is the
limit that the singlet superfield decouples from the MSSM fields; 
the only exception is the effect of its vev $\propto \lambda^{-1}$, 
and in particular, the effective $\mu$ parameter of the MSSM given by 
$\lambda \vev{s}$.
We even find the same familiar minimization constraints on the soft
parameters as in the MSSM. 
One important difference however is the issue of fine tuning, 
since the original parameter space differs from that of the MSSM. 
We find that large $\tan \beta$ is fine tuned.

Since our framework replaces the MSSM superpotential term $\mu H_1 H_2$ 
with $ \lambda S H_1 H_2$, there are two more parameters, 
$m_G^2$ and $m_S^2$, relatively to the MSSM.
However, $m_G^2$ is relevant only to the mass of the Goldstone boson, 
and to nothing else.
Furthermore, our framework has an extra minimization condition for 
the singlet field $s$, and the minimization reveals that small 
$m_S^2$ is favoured in our framework.
Thus, the two extra parameters are eventually irrelevant 
to the effective MSSM, and the extra minimization condition 
leads to an extra prediction in the effective MSSM:
the effective $\mu$ parameter is given by $\lambda \vev{s}$, and 
$\mu \approx A/\tan \beta$. 
Not only is the $\mu$ problem solved, but the $\mu$ parameter is 
predicted \cite{CP}. It will be very important to test this relation, 
which relates the chargino/neutralino masses and the
heavy Higgs scalar masses to the value of $\tan \beta$. Note, however,
that this prediction survives even when $\lambda \approx 10^{-9}$ and
there is no explicit PQ breaking beyond QCD (so that the Goldstone 
becomes the invisible axion) and hence cannot be used to distinguish
our electroweak framework from supersymmetric invisible axion theories 
\cite{MN}.

The production and decays of superpartners and Higgs bosons at particle 
accelerators is very similar to the MSSM, with one crucial difference. 
The LSP of the MSSM is the next-to-LSP (NLSP) of our theory. Hence in 
our theory all the supersymmetric processes end with the NLSP decays 
to the LSP and a standard model particle that is the ``superpartner'' 
of the NLSP. This radically changes the collider signals of supersymmetry, 
and allows regions of parameter space where the NLSP is charged. (In the MSSM
the corresponding particle is the LSP and, if it is stable, cosmology 
requires it to be neutral.) 
The modification of the signal from NLSP decay clearly depends on what 
the NLSP is. Examples of such decays include $\tilde{\chi}^0_1  \rightarrow  
(Z,h) \tilde{\chi}_0$, $\tilde{\chi}^+_1  \rightarrow   W^+ \tilde{\chi}_0$,
$\tilde{t}  \rightarrow   t \tilde{\chi}_0$ and
$\tilde{\tau} \rightarrow  \tau \tilde{\chi}_0$, so that superpartner
pair production will lead to events with pairs of $(Z,h), W, t$ and
$\tau$ respectively. In each of these cases the 
missing (transverse) energy is degraded. The MSSM signals would be unchanged 
only if the NLSP is the sneutrino, with
$\tilde{\nu}  \rightarrow  \nu \tilde{\chi}_0$.

The light $G$ and $\tilde{\chi}_0$ states could play an important
role in cosmology. The LSP  $\tilde{\chi}_0$ could be the dark
matter of the universe, providing there is a large amount of entropy
generated in the universe well after the electroweak scale but before
BBN. If $m_G < T_\nu \sim 3$ MeV, then one must include either $G$
or its decay products in the calculations of the effective number of
neutrino species at BBN and CMB eras. For the CMB case we find
$N_{\nu CMB} = 2.4$, while the result for $N_{\nu BBN}$ is sensitive to
$m_G$ and could apparently be slightly above or below the usual value
of 3.

There is only a narrow window left for $\vev{s}$, 
from $10^2$ TeV to $10^3$ TeV, for small pseudo-Goldstone boson mass.
Thus, it is important to search for $G \rightarrow e^+ e^-$ decay at 
improved reactor experiments, and to search for more 
$K^+ \rightarrow G +\pi^+$ events at kaon factories.

\section*{Acknowledgments}
\label{acknowledgements}

This work was supported in part by the Miller Institute 
for the Basic Research in Science (T.W.), by the Director, 
Office of Science, Office of High Energy and Nuclear Physics, of the
U.S. Department of Energy under Contract DE-AC03-76SF00098 
and DE-FG03-91ER-40676, and in
part by the National Science Foundation under grant PHY-00-98840.

\end{document}